\documentclass[twocolumn, twocolappendix, numberedappendix]{openjournal}

\usepackage{xcolor}
\usepackage{textgreek}
\usepackage[utf8]{inputenc}
\usepackage[english]{babel}

\usepackage{hyperref}
\hypersetup{
    unicode, 
    colorlinks=true,
    linkcolor=linkcolor,
    citecolor=linkcolor,
    filecolor=linkcolor,
    urlcolor=linkcolor,
}
\usepackage{color,colortbl}
\definecolor{linkcolor}{rgb}{0.0,0.3,0.5}
\usepackage{tensind}
\tensordelimiter{?}
\DeclareGraphicsExtensions{.bmp,.png,.jpg,.pdf}
\usepackage{verbatim}
\usepackage[normalem]{ulem}
\usepackage{orcidlink}
\usepackage{soul}

\graphicspath{ {./figs/} }

\shorttitle{Blending Effects in LSST Clustering}
\shortauthors{Levine, S\'{a}nchez et al.}

\begin{document}
\title[Blending Effects in LSST Clustering]{Galaxy Clustering with LSST: Effects of Number Count Bias from Blending}

\author{\vspace{-4em}Benjamin Levine,$^{\ast,1,2}$
Javier S\'{a}nchez,$^{\dagger,3,4,5}$
Chihway Chang,$^{2,4}$
Anja von der Linden,$^{1}$\\
Eboni Collins,$^{6}$
Eric Gawiser,$^{7}$
Katarzyna Krzy\.za\'{n}ska,$^{8}$
Boris Leistedt,$^{9}$\\
and The LSST Dark Energy Science Collaboration}

\affiliation{\vspace{-0.7em}}
\affiliation{$^{1}$Department of Physics and Astronomy, Stony Brook University, Stony Brook, NY 11794, USA}
\affiliation{$^{2}$Department of Astronomy and Astrophysics, University of Chicago, Chicago, IL 60637, USA}
\affiliation{$^{3}$Space Telescope Science Institute, 3700 San Martin Dr., Baltimore, MD 21218, USA}
\affiliation{$^{4}$Kavli Institute for Cosmological Physics, University of Chicago, Chicago, IL 60637, USA}
\affiliation{$^{5}$Fermi National Accelerator Laboratory, P.O. Box 500, Batavia, IL 60510, USA}
\affiliation{$^{6}$Department of Physics, Dillard University, 2601 Gentilly Blvd, New Orleans, LA 70122, USA}
\affiliation{$^{7}$Department of Physics and Astronomy, Rutgers University, Piscataway, NJ 08854, USA}
\affiliation{$^{8}$Department of Physics, Cornell University, 109 Clark Hall, Ithaca, NY 14853, USA}
\affiliation{$^{9}$Department of Physics, Imperial College London, Blackett Laboratory, Prince Consort Road, London SW7 2AZ, UK}

\email[$^\ast$]{benjamin.c.levine@stonybrook.edu}
\email[$^\dagger$]{jsanchez@stsci.edu}

\begin{abstract}
    The Vera C. Rubin Observatory Legacy Survey of Space and Time (LSST) will survey the southern sky to create the largest galaxy catalog to date, and its statistical power demands an improved understanding of systematic effects such as source overlaps, also known as blending. In this work we study how blending introduces a bias in the \textit{number counts} of galaxies (instead of the flux and colors), and how it propagates into galaxy clustering statistics. We use the $300$\,deg$^2$ DC2 image simulation and its resulting galaxy catalog \citep{2020arXiv201005926L} to carry out this study. We find that, for a LSST Year 1 (Y1)-like cosmological analyses, the number count bias due to blending leads to small but statistically significant differences in mean redshift measurements when comparing an observed sample to an unblended calibration sample. In the two-point correlation function, blending causes differences greater than 3$\sigma$ on scales below approximately $10'$, but large scales are unaffected. We fit $\Omega_{\rm m}$ and linear galaxy bias in a Bayesian cosmological analysis and find that the recovered parameters from this limited area sample, with the LSST Y1 scale cuts, are largely unaffected by blending. Our main results hold when considering photometric redshift and a LSST Year 5 (Y5)-like sample.
\end{abstract}


\maketitle

\section{Introduction}
\label{sec:intro}

The Vera C. Rubin Observatory Legacy Survey of Space and Time \citep[LSST,][]{2019ApJ...873..111I} is projected to begin its 10-year survey of the southern hemisphere sky in late 2025. Both the depth and sky coverage of LSST are significantly enhanced with respect to Stage-III dark energy experiments~\citep{2006astro.ph..9591A} such as the Dark Energy Survey \citep[DES,][]{Flaugher2005}, the Kilo-Degree Survey \citep[KiDS,][]{DeJong2015} and the Hyper Suprime-Cam Subaru Strategic Program \citep[HSC-SSP,][]{Aihara2017}. These improvements are expected to lead to the detection of approximately $20$ billion galaxies, improving the statistical power from a variety of dark energy probes~\citep[see the LSST Dark Energy Science Collaboration Science Requirement Document,][hereafter ``DESC SRD'']{2018arXiv180901669T}. However, the improved statistical power will also lead to an increased sensitivity to systematics as previously-subdominant effects become relevant. 

One systematic effect of particular interest to static probes of cosmology is blending systematics, an effect in which the detection and/or measurement algorithms mischaracterize—or fail to detect entirely—overlapping luminous sources. This effect is especially prominent in faint galaxy populations, as the number density of objects increases steeply with fainter magnitude, meaning that the likelihood of overlapping another source with similar brightness also increases steeply. Blending can bias the centroid position, color, and shape of sources—all quantities that are used in cosmological probes of the large-scale structure~\citep[e.g.,][]{2018PhRvD..98d3526A, KiDS1000_cosmo}. In the HSC-SSP, more than 50\,\% of detected objects were ``deblended'' \citep{Bosch2018}; in image simulations of a joint Roman Space Telescope and Rubin Observatory survey, 20\,\%--30\,\% of Rubin single detections were identified as multiple blended objects in Roman data \citep{Troxel_2023}. For a more detailed review of blending in cosmic surveys, see~\cite{Melchior2021}.

The effects of blending in the context of weak lensing analyses have seen interest from the community in recent years~\citep[e.g.,][]{2016ApJ...816...11D, 2017ApJ...841...24S,  2019MNRAS.488.4389G,  2021arXiv210302078S, Nourbakhsh2022}. Various techniques have been developed to minimize the impact to cosmic shear estimation~\citep[e.g.,][]{2016MNRAS.459.4467B, 2020ApJ...902..138S} and photometric redshifts~\citep[e.g.,][]{2019MNRAS.483.2487J, MacCrann_2021}. New deblending techniques are also being developed in order to mitigate the effects of blending in different probes~\citep[e.g.,][]{2015A&A...582A..15M, 2016A&A...589A...2J, 2018A&C....24..129M, 2021MNRAS.500..531A}. However, the impact of blending in the study of two-point statistics of galaxy positions (i.e., galaxy clustering) has not received such attention. This lack of interest is partially due to the fact that blending is mostly thought to be important on very small scales where the separation between galaxies is comparable to the size of the galaxies themselves, and these spatial scales are usually not used in cosmological analyses. In addition, the effect is not seen in most current datasets due to the lower galaxy number density. However, as we approach the era of LSST, it is important to revisit the question of whether and how blending could affect galaxy clustering measurements in Stage-IV surveys. 

In principle, blending will have three main effects on galaxy clustering measurements: 1) Unrecognized blends cause us to “lose” galaxies from the catalog; we thus count fewer pairs on the scales at which blending occurs, as well as the characteristic clustering scales of galaxies that cause blending and galaxies that are susceptible to blending. 2) Blending can cause errors in the brightness measurements of a galaxy, causing it to scatter into or out of a magnitude-selected sample. 3) Blending can cause errors in the color measurements of a galaxy, resulting in incorrect photometric redshift estimation.

The third effect is both complex to measure and highly algorithm-dependent \citep{MacCrann_2021}; in addition, characterizing it would require very high accuracy and precision in simulating the galaxy population, which is hard to achieve. There are, however, ongoing efforts to produce reliable photometric redshifts for LSST \citep[e.g.,][]{2020MNRAS.499.1587S}. In Appendix \ref{app:results_photz} we explore a photometric redshift scenario using Bayesian Photometric Redshift~\citep[BPZ,][]{BPZ2000} redshift estimates from simulated photometry. However, due to the aforementioned complications, we leave a thorough examination of photometric redshift impacts for future study.

The first two effects are easier to assess and are the focus of this work. The LSST Data Challenge 2~\citep[DC2,][]{2020arXiv201005926L} image simulations allow us to implicitly capture both effects by comparing the ``observed''\footnote{Throughout this work we will refer to the simulated DC2 photometric catalog as ``observed'' data, since it corresponds directly to what we expect to observe with LSST.} DC2 Year 1 (Y1) correlation function to the corresponding correlation function in the truth catalog, without needing to quantify some of the details (e.g., measuring how much the magnitudes are typically biased, or deciding how to identify a blend in detail). Then, to track down the specific effects of blending on galaxy clustering (as opposed to any other effects introduced by the image processing pipeline), we choose a relatively simple definition to approximately distinguish between blended and unblended objects.

We emphasize here that, although we will spend considerable time describing our selected definition of blended and unblended objects, many of our results—in particular those concerning any shifts in the cosmology inference—are independent of this definition because they rely only on a comparison between the truth and observed catalogs. The definition of blends serves mainly to distinguish between biases that arise from blending and biases that arise from some other source.

We focus particularly on the effect on the redshift distribution and inferred values of matter density $\Omega_{\rm m}$ and linear galaxy bias $b(\bar{z})$. We will not investigate any additional related issues (including color shifts, shape measurements, and centroid offsets), but future studies could in principle adopt the same tools and methodology to analyze them.

This work is structured as follows: in Section~\ref{sec:methods} we present the DC2 samples used for our analysis, as well as the methodology followed. In Section~\ref{sec:z_results} we show the results regarding the bias to the redshift distributions of our samples. In Section~\ref{sec:w_results} we present the clustering results obtained for the different samples considered in this work. In Section~\ref{sec:limitations} we discuss the limitations of our analysis. Finally, in Section~\ref{sec:conclusions} we present concluding remarks.

\section{Analysis Framework}
\label{sec:methods}

\subsection{Data}
\label{ssec:data}

\begin{figure*}
\centering
    \includegraphics[width=\textwidth]{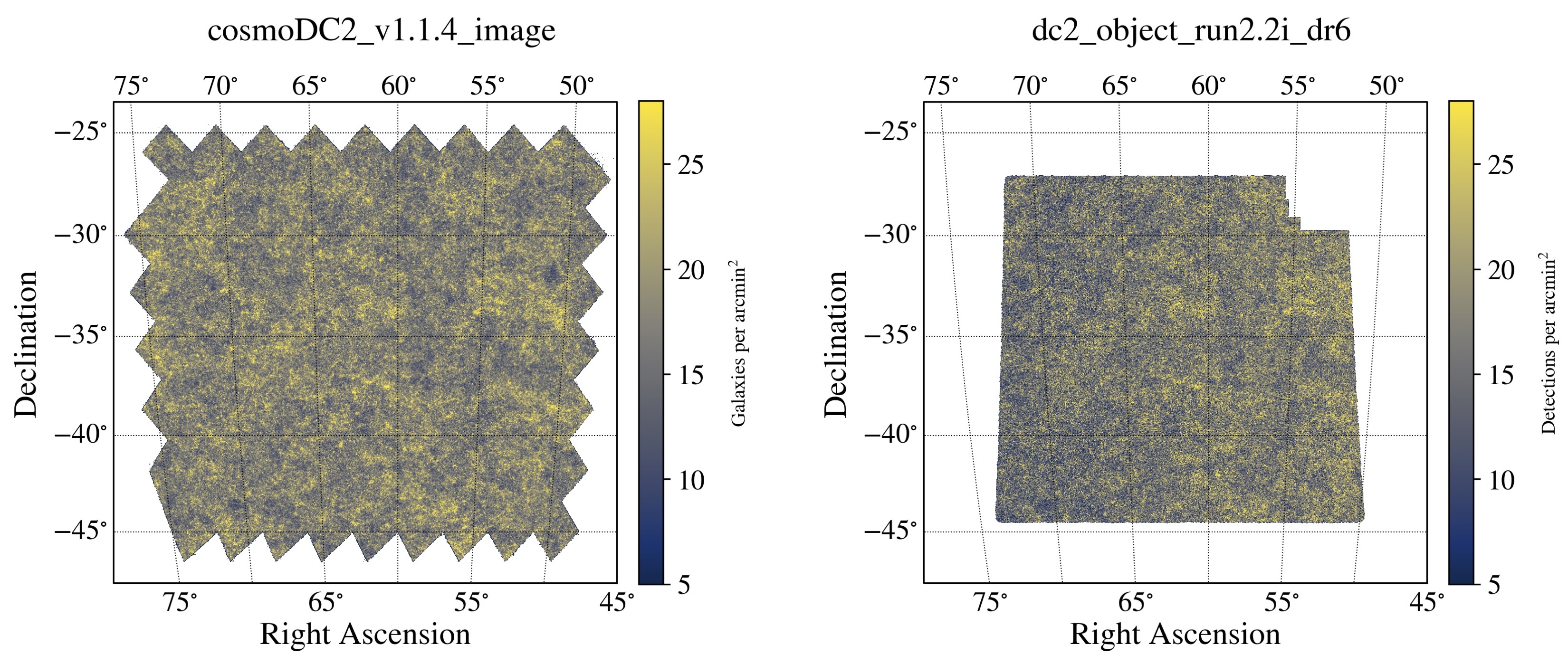}
    \caption{\emph{Left}: map of the input (truth) galaxies from the CosmoDC2 simulation. \emph{Right}: map of detected sources in the DC2 DR6 object catalog. The maps are shown using the McBryde projection.}
    \label{fig:footprint}
\end{figure*}

It is challenging to infer the impact of blends in LSST using existing survey data: some existing ground-based datasets are deep enough to achieve the required number densities to evaluate the impact of blending (such as the HSC-SSP deep and ultra-deep fields), but their small area is inadequate to test the full cosmological parameter space to the precision required by LSST. Using larger but shallower data (such as HSC-SSP wide) does not guarantee mapping the full parameter space relevant to blending (e.g., morphology, size, surface brightness, and clustering amplitude). One solution is to supplement shallower surveys with overlapping data from deeper, better-resolved images, e.g., with existing space-based imaging from HST; or very large telescopes with adaptive-optics systems. However, this approach quickly becomes limited by the expense of telescope time given the large number of galaxies we wish to study. In addition, there is no guarantee that there are no sources below the noise level that contribute to the measured flux of our detected sources. Future space-based surveys such as Roman~\citep{Roman2015, Roman2019} and Euclid~\citep{Euclid2011} will provide the required data, but their data products will not be available in time to prepare for LSST. To this end, realistic end-to-end image simulations offer a unique opportunity to study the blending problem because we have access to both the catalog produced by our detection and deblending algorithms and, most importantly, the truth. In particular, the DESC DC2 simulations are developed specifically for LSST. Despite their limited area (approximately $300$\,deg$^2$) compared to the full planned area for LSST (approximately $18000$\,deg$^2$), the access to ground truth and their underlying number density makes them extremely useful tools with which to study the effects of blending.

DC2 is an end-to-end photometric catalog based on the Outer Rim simulation~\citep{2019ApJS..245...16H}. It includes a 440\,deg$^{2}$ cosmological mock galaxy catalog (CosmoDC2) and a synthetic stellar catalog, simulated single-epoch and coadd images and their corresponding catalogs in six bands ($u, g, r, i, z, y$), and a combined forced-photometry catalog. A detailed description of the different data products from DC2 can be found in~\citet{2020arXiv201005926L}. The CosmoDC2 truth catalog is complete to a magnitude of $r \sim 28$ and up to redshift $z = 3$. It also includes a large population of faint galaxies up to $r \sim 33$ to induce realistic blending effects. The galaxies are generated using a mixture of a semi-analytical model (SAM) based on Galacticus~\citep{2012NewA...17..175B} and an empirical model based on UniverseMachine~\citep{2019MNRAS.488.3143B}. The simulation has clustering properties consistent with a flat $\Lambda$CDM universe, with cosmological parameters closest to the WMAP-7 best-fit~\citep{2011ApJS..192...18K}: $\omega_{cdm} = 0.1109,\, \omega_{b} = 0.02258,\, n_{s}=0.963,\, h=0.71,\, \sigma_{8}=0.8$. Galaxy morphologies, shapes, and sizes are obtained from the SAM and consist of combinations of S\'{e}rsic bulge ($n=4$) and disk ($n=1$) models. Bright sources include a random walk component to mimic galaxy knots. Additional details about these and other properties of CosmoDC2 are summarized in~\citet{2019ApJS..245...26K}, and~\citet{2020arXiv201005926L}.

We focus on the Wide-Fast-Deep (WFD) area shown in Figure~\ref{fig:footprint}. We use the \texttt{dc2\_run2.2i\_object\_dr6} 5-year LSST depth ``object catalog,'' produced by detecting sources in the coadded images, for our observed data. Using this 5-year depth catalog guarantees a high completeness for the Y1-like subsample that we analyze here, minimizing potential impacts of incompleteness that may dominate over blending effects. This catalog is processed with version 19.0.0 of the Rubin Science Pipelines\footnote{For additional details on the Rubin Science Pipelines see \url{pipelines.lsst.io}.} \citep{RubinPipelines2015}, and is designed to resemble the catalogs that will be produced during the operations of LSST. Detailed validation tests were carried out on the DC2 products as described in~\citet{Kovacs2022}, and additional validation is shown in Appendix \ref{app:skysim}. To avoid contamination and systematic impacts on the sample by bright stars, we apply the star mask developed by \cite{Du_2023}. Excluding this mask does not significantly affect our results.

The end-to-end simulation framework of DC2 offers us the unique opportunity to study blending in a realistic yet controlled way: we can match the galaxy catalog as detected by the Rubin software to the input ``true'' galaxy sample, in principle allowing us to determine the effects of blending on any given object. It is important to note that some of the effects of blending are tied to the choice of detection and deblending algorithm used for the analysis.\footnote{Additional details about the detection and deblending algorithms that we use can be found in~\citet{Bosch2018}.} When we measure properties based on the truth catalog, such as the intrinsic redshift distribution of isolated vs.\ blended objects (Section \ref{sec:z_results}), any observed biases will be present independently of the detection and deblending algorithms, since our definition of isolated and blended objects is independent of these algorithms. However, for properties based on derived quantities in the observed catalog, such as flux and position (both of which propagate into the two-point correlation function; Section \ref{sec:w_results}), the bias will be dependent on the specifics of the detection and deblending algorithms.

\subsection{Matching truth and observed catalogs}
\label{ssec:matching}
The information of which photon was emitted by which source is not propagated through to the simulated images, and therefore there is no straightforward way to determine which truth objects were blended together into a single detection. As a consequence, we match objects in the observed catalog to objects in the truth catalog by selecting the truth object closest in $r$-band magnitude within a given radius $R_{\rm{max}}$, making the assumption that only objects within $R_{\rm{max}}$ will likely be blended. We match in the $r$ band because it is the deepest photometric band. The radius size $R_{\rm{max}}=1''$ is chosen to be approximately the size of the typical point-spread function (PSF) of LSST images \citep[estimated to be approximately $0.7''$ for the $r$ band,][]{2019ApJ...873..111I}. We select our galaxy sample to have an observed magnitude of $i<24.1$ (which follows the LSST Y1 Gold sample magnitude cut as specified in the DESC SRD) and to be extended in each one of the \textit{g, r, i,} and \textit{z} bands (\texttt{<band>\_base\_ClassificationExtendedness\_value} $> 0.5$). Due to this second requirement, some compact galaxies may be misclassified as stars and be incorrectly removed from the sample. The number of removed galaxies is $1.4\,\%$ of the total number of observed galaxies in our main analysis. We also check the number of stars that are misclassified as galaxies: these objects make up $1.2\,\%$ of our observed sample. However, our fiducial analysis involves tomographic binning on the true redshift of objects, which will remove all of these misclassified stars. Star-galaxy misclassifications are therefore unlikely to introduce a bias to our results.

\begin{figure}
    \includegraphics[width=\columnwidth]{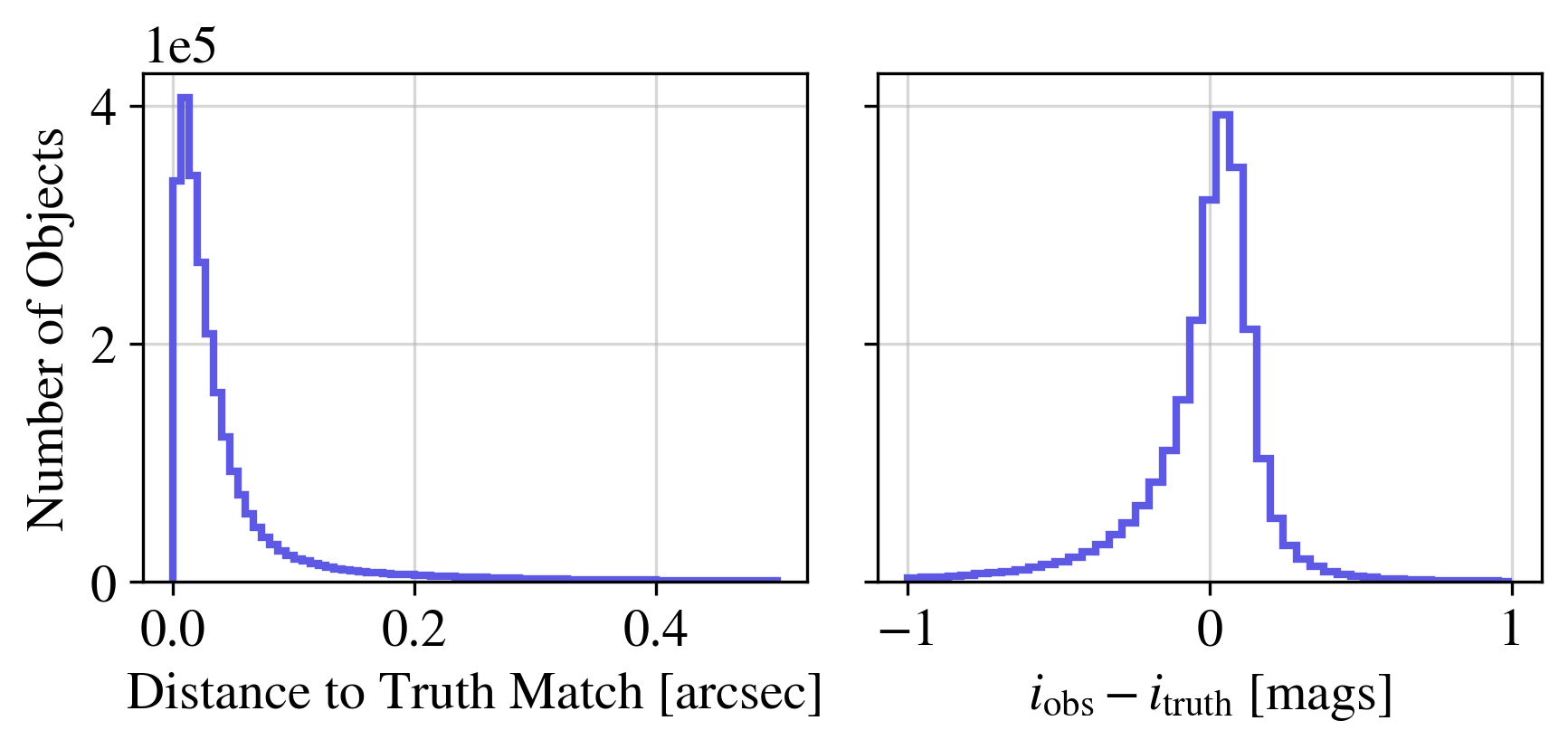}
    \caption{\emph{Left}: histogram of the distance between an observed object and its best truth match. \emph{Right}: histogram of the difference between the $i$-band magnitude of an observed object and its best truth match. For the right plot we use objects from \texttt{cosmoDC2\_v1.1.4\_small}, a representative subset of CosmoDC2. For both, the best match is defined as the truth object nearest in $r$-band magnitude to the detection within $1''$.}
    \label{fig:distance}
\end{figure}

Figure \ref{fig:distance} shows that nearly 100\,\% of our selected observed objects in DC2 are within $0.4''$ from their best truth match, suggesting that the pipeline adequately distinguishes between centroid separations greater than $1''$—if not, we would expect to see more matches at larger radii. $R_{\rm{max}}=1''$ is therefore a conservative choice of matching radius. Figure \ref{fig:distance} also shows that over 95\,\% of our observed objects are within 0.25 magnitudes of their best match in the truth catalog, providing further evidence for the reliability of our matching process.

\begin{figure}
    \includegraphics[width=\columnwidth]{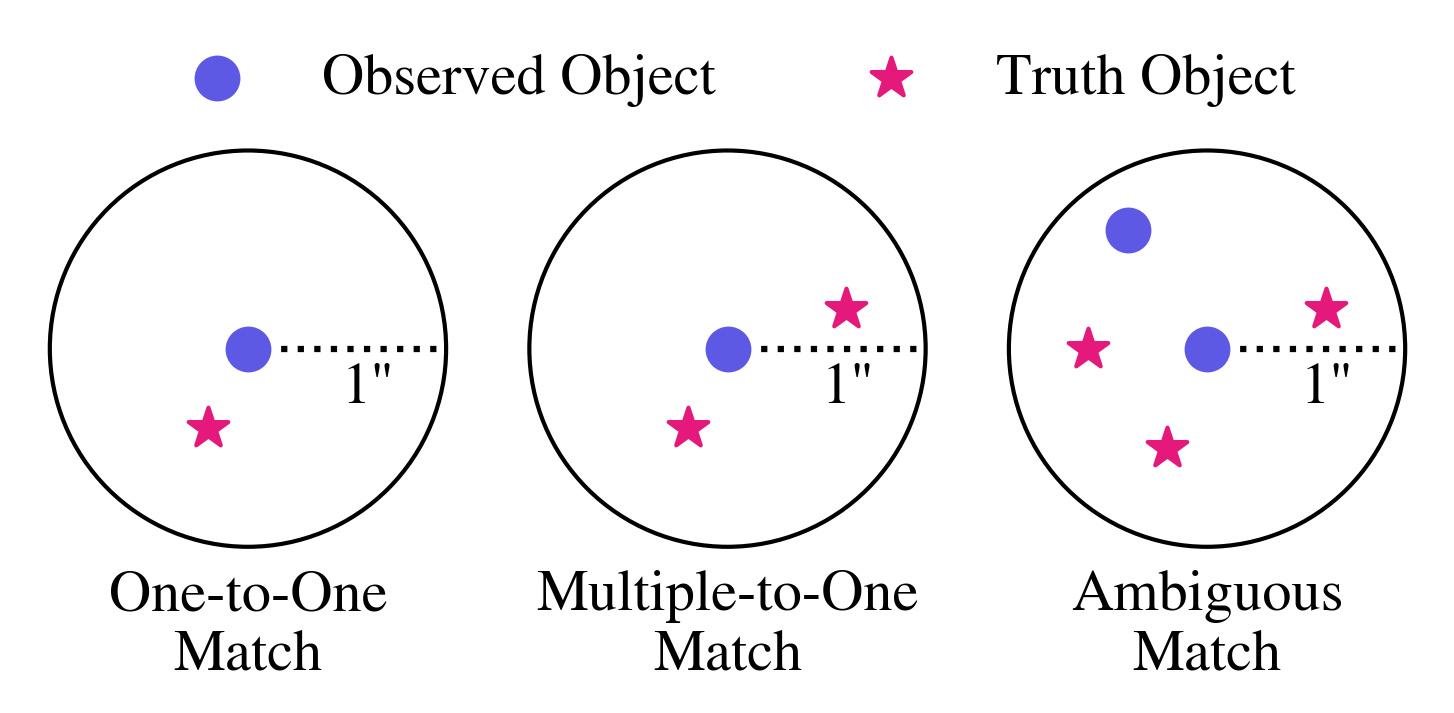}
    \caption{The selection criteria for matching objects between the observed and truth catalogs. We perform a search within $1''$ of the observed object. One-to-one matches (left) are defined as $N\textsubscript{truth} = N\textsubscript{observed} = 1$ and multiple-to-one matches (center) are defined as $N\textsubscript{truth} > N\textsubscript{observed} = 1$. Ambiguous matches (right) are defined as $N\textsubscript{observed} > 1$.}
    \label{fig:sample_selection_descript}
\end{figure}

We use a KDTree algorithm from \textsc{scikit-learn}~\citep{scikit-learn} to count the number of neighbors within $R_{\rm{max}}$ in the truth and observed catalogs for each observed detection. Our definition of neighbor includes the centered object; i.e., an isolated observed object with no other detections within $R_{\rm{max}}$ will have one observed neighbor (itself). Then, we sort the matches into three broad categories (see Figure \ref{fig:sample_selection_descript} for an illustration).

\begin{itemize}
    \item \textbf{One-to-one matches:} all sources in the observed catalog that have only one neighboring source in the truth catalog within a radius $R_{\rm{max}}=1''$, and one neighboring source within $R_{\rm{max}}=1''$ in the observed catalog. These sources will most likely be unaffected by flux from any other objects (i.e., we can consider them isolated). The tails of very bright sources could in principle extend beyond $R_{\rm{max}}$ and influence the determination of other sources of interest \citep{Gawiser2006}. However, we expect that the majority of the sources impacted by these effects will be masked by the bright star mask and thus excluded from the sample. Approximately 42\,\% of observed galaxies in our sample are one-to-one matches. 
    \item \textbf{Multiple-to-one matches:} all sources in the observed catalog that have more than one neighboring source within $R_{\rm{max}}=1''$ in the truth catalog and one neighboring source within $R_{\rm{max}}=1''$ in the observed catalog. The photometry and astrometry of these sources could be biased and the uncertainties may be correlated with neighboring sources. These detections are considered to be blended. Approximately 57\,\% of observed galaxies in our sample are multiple-to-one matches. 
    \item \textbf{Ambiguous matches:} all sources in the observed catalog which have more than one neighbor in the observed catalog. This category may also include shreds, in which a single true object may lead to multiple detections in the observed catalog. We leave the analysis of this category to future work.\footnote{Note that the nomenclature for categories of blends is not standardized. The so-called ``ambiguous blends'' described in \cite{2016ApJ...816...11D} correspond to our multiple-to-one matches; \cite{Mandelbaum_2018} uses ``unrecognized blends'' to refer to the same.} Given their rarity (approximately 1\,\% of our sample) we do not expect ambiguous blends to significantly affect the results of this study. In addition, an analysis of this category would require a more sophisticated matching algorithm that takes into account the physical overlap of the measured shapes of objects; such algorithms are currently in development (e.g., \citealt{Ramel:2023fgr}; Liang, Adari et al. in prep.).
\end{itemize}

We count all neighbors in the truth and observed catalogs, up to a magnitude $i<30$. In practice, there will not be significant bias to a galaxy blended with an object many magnitudes fainter than itself. However, determining what magnitude cutoff should be used to identify a ``significant'' or ``strong'' blend is nontrivial. Rather, the work here should be interpreted as the worst-case scenario—i.e., that almost every blended galaxy has unreliable photometry. We also reiterate that the exact definition of our one-to-one and multiple-to-one samples serves mainly to identify which biases are related to blending and which are not. The majority of our results are independent of this definition.

A benefit of using these distance-based categories compared to other criteria based, for example, on the degree of blending~\citep[such as blendedness;][]{Bosch2018} is that they do not depend as strongly on the choice of detection and deblending algorithms. 

It should be noted that it is extremely challenging to identify these three categories in real (unsimulated) data (though, as we will discuss in Section \ref{sec:z_results}, a subsample of bright, isolated galaxies, such as those used for spectroscopic redshift calibration, should be close to a one-to-one sample). To emulate the expected LSST Y1 observations, we also present results for the ``all observed'' category, which consists of all galaxies detected in the DC2 observed catalog (the union of the three categories of blending described above).

We introduce a final category, defining ``lost'' objects as truth objects that are not a nearest match for a detection in the observed catalog. Note that the objects in this category do depend on the particular choice of detection and deblending algorithms.

\begin{figure}
    \includegraphics[width=\columnwidth]{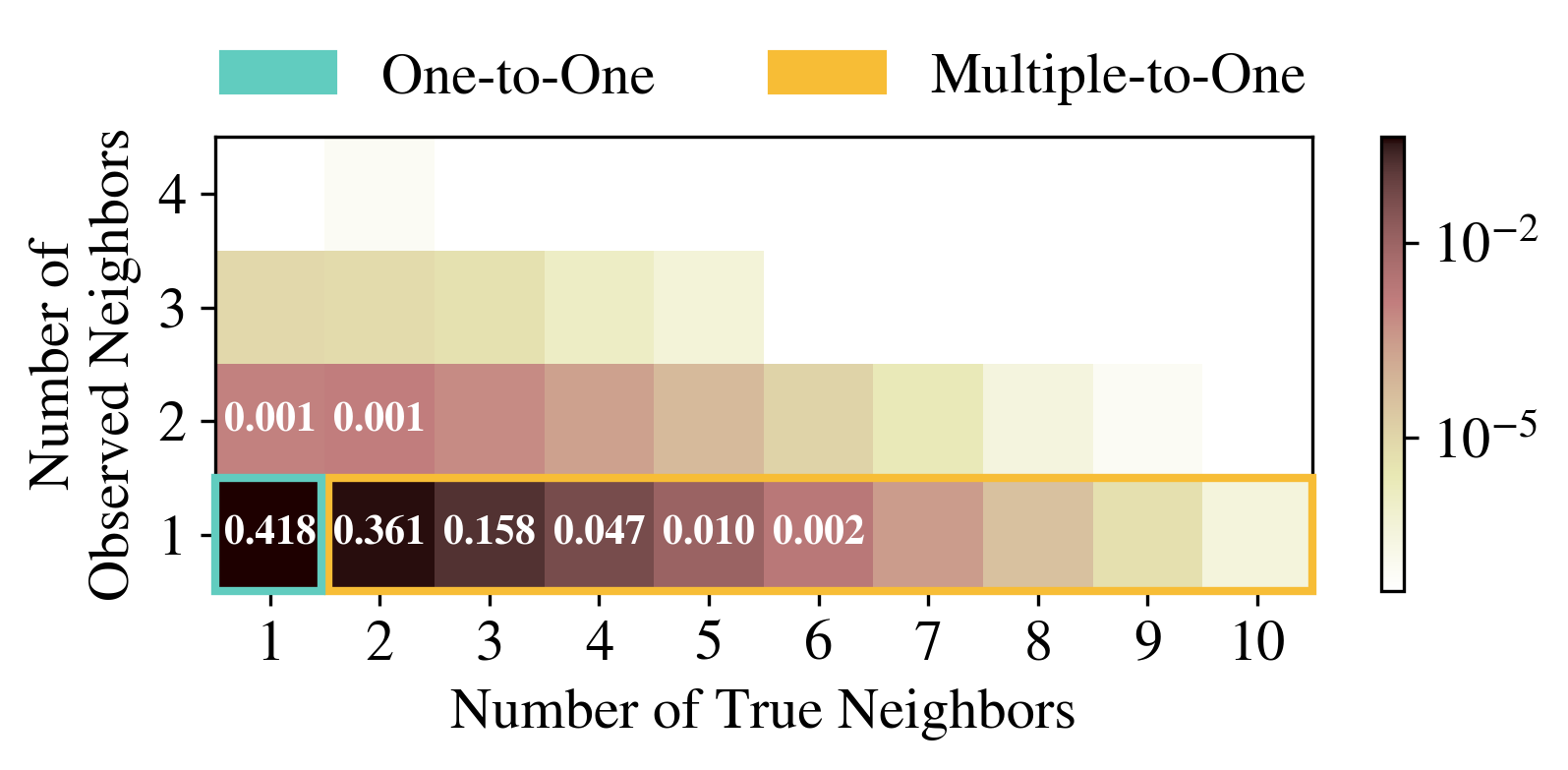}
    \caption{Distribution of neighbors in the truth and observed catalogs for each detected object, with a magnitude cut of $i<24.1$. Cells corresponding to our one-to-one and multiple-to-one samples are highlighted.}
    \label{fig:blends_proportions}
\end{figure}

In Figure~\ref{fig:blends_proportions} we show the distribution of observed sources as a function of the number of neighbors in the truth and observed catalogs. The histogram cells above the diagonal are spurious detections due to source shredding or poor pixel masking. The cells on the diagonal are sources that have been ``correctly'' detected, i.e., there are no biases between the input and output number of sources. The cells below the diagonal correspond to sources that have undetected neighbors. Approximately 42\,\% of the sources are one-to-one matches, i.e., are most likely not blended, and approximately 57\,\% of the detected sources have more than one truth neighbor within $1''$, meaning that these sources likely present some degree of blending. This means that more than half of the detected objects in DC2 Data Release 6 (DR6) present some degree of blending~\citep{Melchior2021}.

Note that this fraction is substantially higher than that found by \cite{Troxel_2023}, which is likely due the fact that we match to fainter objects and have a different definition of blends. Indeed, when we cut our truth catalog at $i<27$ before matching, we find that only $17.5\,\%$ of objects exhibit blending. Nonetheless, note that Figure 10 in \cite{Troxel_2023} is qualitatively similar to our Figure \ref{fig:blends_proportions}.

\begin{figure}
    \includegraphics[width=\columnwidth]{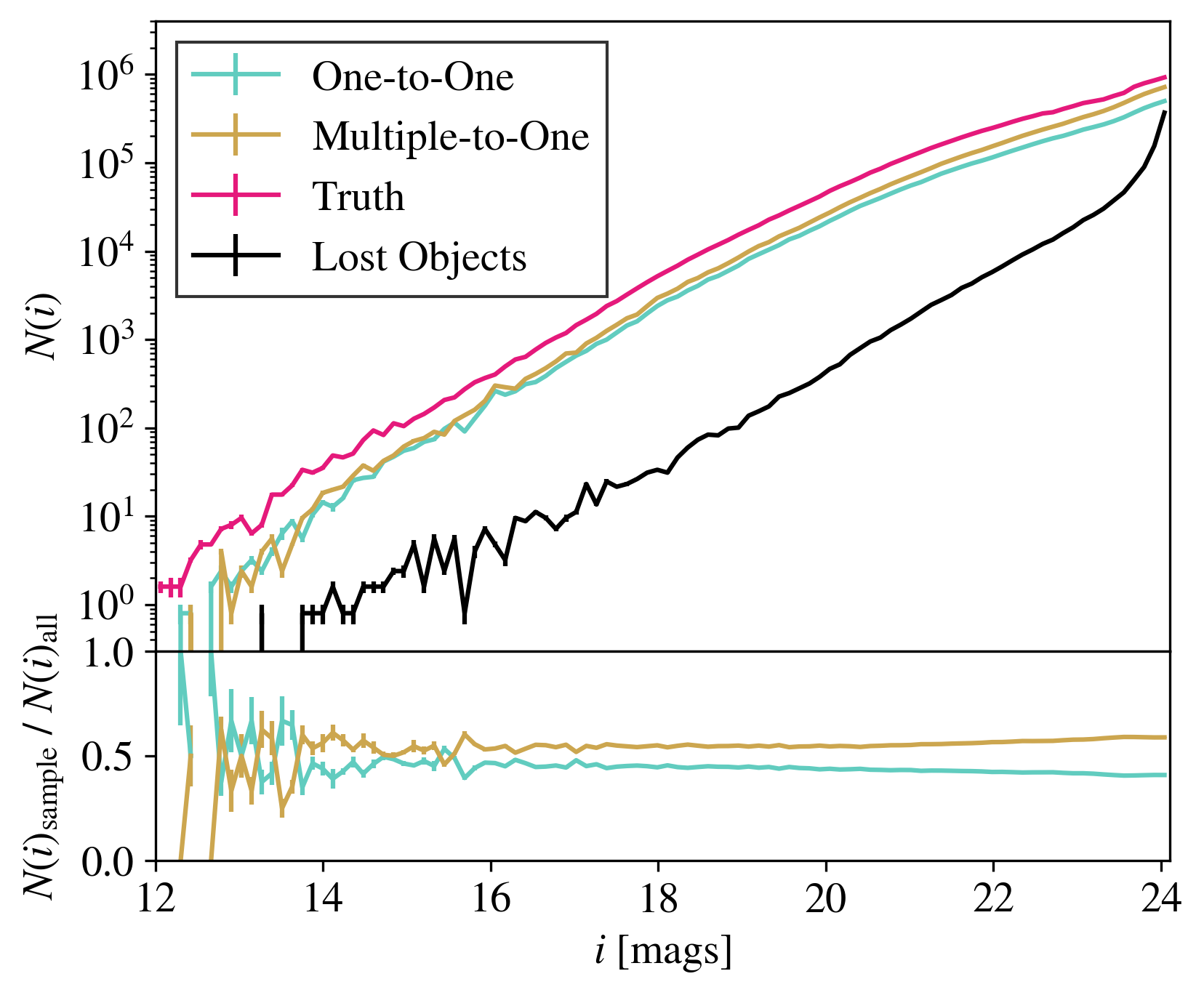}
    \caption{\emph{Top}: number of objects in our samples at a given $i$-band magnitude. For the observed samples, we plot the observed magnitudes derived from the Rubin Science Pipelines. For the truth and lost samples, we plot the true magnitudes from the CosmoDC2 simulation. \emph{Bottom}: deviation of the one-to-one and multiple-to-one samples from the all observed sample.}
    \label{fig:mag_distribution}
\end{figure}

In Figure~\ref{fig:mag_distribution} we show the $i$-band magnitude distribution of the one-to-one matches, multiple-to-one matches, and lost samples. For reference, we also plot the magnitude distribution of the truth catalog. Both detected samples closely follow the true distribution, while the lost objects remain comparatively-rare up to $i = 23$. Fainter objects are more difficult to detect, even if isolated, and are more sensitive to the presence of nearby luminous sources. 

\section{Bias in Redshift Distribution}
\label{sec:z_results}

Cosmological analyses that rely on properties of ensembles of galaxies, such as galaxy clustering and weak lensing shear, require an accurate description of the overall redshift distribution of the sample of galaxies involved. In this section, we test whether blending affects the calibration of these distributions using reference spectroscopic samples in a measurable way. 

Although it is difficult to identify blended objects in a full cosmology catalog, there are circumstances in which it is possible to identify a likely-unblended sample. The DIR redshift calibration method~\citep{2008MNRAS.390..118L}, for example, relies on a representative sample of spectroscopic redshifts to characterize the $N(z)$ distribution of the full sample. Spectroscopy is typically performed on bright, isolated galaxies, meaning that these calibration samples are likely to be unblended. If the underlying redshift distribution of unblended objects is different from that of blended objects, this calibration method will introduce a bias to the measured $N(z)$.

We limit our analysis to testing for changes in the ensemble redshift distributions knowing the underlying truth redshifts and colors. However, blending can affect photometric redshift calibration in other ways (e.g., using unrecognized blends in the calibration sample, which can lead to biased redshift estimates). These effects, which are dependent on the calibration method used, are left for future work.

\begin{figure}
    \includegraphics[width=\columnwidth]{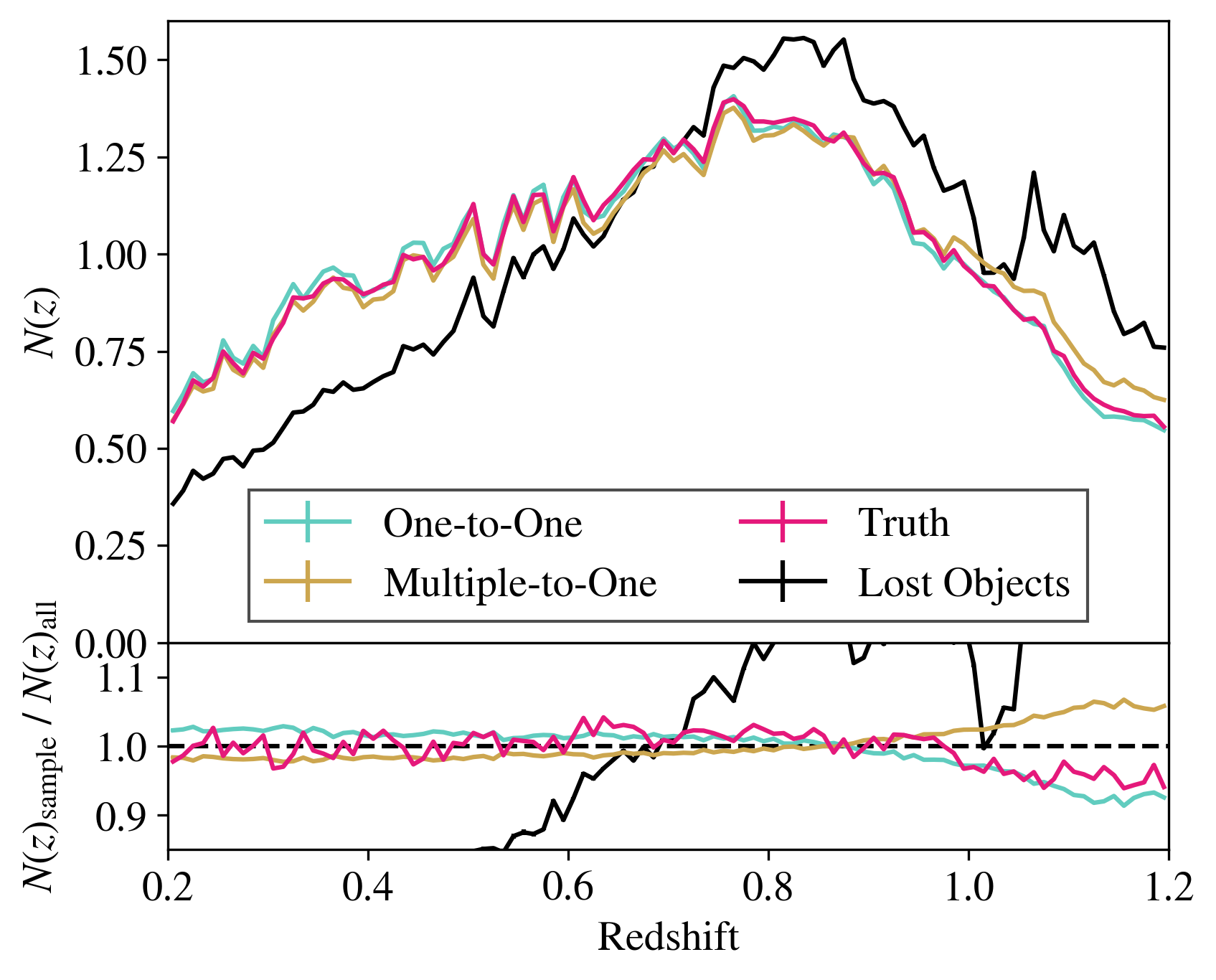}
    \caption{\emph{Top}: number of objects in our samples at a given redshift, normalized such that the integral over all bins is equal to unity. For the observed samples, the redshift is given by the true redshift of the best truth match. \emph{Bottom}: deviation from the all observed sample.}
    \label{fig:z_dist_all}
\end{figure}

Figure \ref{fig:z_dist_all} shows the probability density of each of our samples as a function of redshift. For the observed catalog, we use the true redshift of the best match in the truth catalog (as described in Section \ref{ssec:matching}). The redshift of the multiple-to-one and lost samples are skewed to higher redshifts (to an extreme degree for the lost sample) compared to the one-to-one, all observed, and truth samples; i.e., blending tends to affect objects at higher redshifts. Given our definition of blends, one would naively expect blending to occur independently of both redshift and magnitude, since neither the redshift nor magnitude of objects ever enter into our $1''$ matching scheme. There are two factors which contribute to the observed bias. First, at higher redshifts, the angular diameter distance becomes smaller, meaning that physical clustering within a halo occurs at smaller angular scales. When the separation between halo members becomes smaller than $1''$, these high redshift galaxies will be identified as blends by our algorithm. Second, blending increases the measured flux of dim objects close to our magnitude cutoff, causing objects that would otherwise be excluded from our sample to be falsely included. Since these faint objects tend to lie at higher redshifts, we expect the high redshift end of our sample to be preferentially blended. These effects combined qualitatively explain the effect observed, though we will not attempt any explicit forward-modeling in this paper.

The mean and median redshifts of each observed distribution are shown in Table~\ref{tab:table1}. The uncertainties are computed via jackknife with 10 bins, and are order $\mathcal{O}(10^{-6})$—much smaller than the errors on the photometric redshifts that will be used in the LSST cosmological analysis. The measurements show small but significant differences between each sample, with the multiple-to-one sample having the highest mean redshift. 

One consequence of this difference is that redshift calibration methods that rely on the characterization of isolated (one-to-one) galaxies may produce biased estimates of $N(z)$ in the presence of blends. Indeed, Figure \ref{fig:z_dist_all} demonstrates that a one-to-one calibration sample overestimates the normalized $N(z)$ of the all observed sample by 2.27\,\% for $z<0.4$, and underestimates $N(z)$ by 5.92\,\% for $z>1.0$. These biases are statistically significant in DC2, and therefore will also be significant at the statistical level of LSST when trying to characterize the whole ensemble of galaxies at once. This means that the inclusion of nuisance parameters to allow certain perturbations in shape or mean of the $N(z)$, and/or alternative methods to cross-check the overall $N(z)$ will become even more important~\citep[e.g.,][among others]{Cawthon2022, GarciaGarcia2023, RuizZapatero2023}. 

We subdivide our samples into 5 different redshift top-hat bins with 0.2 bin width, between $z=0.2$ and $z=1.2$, to estimate the effects on a tomographic analysis. This follows the Y1 analysis choices described in the DESC SRD. The mean redshifts for each bin, as well as the differences between the one-to-one and all observed samples, are shown in Table~\ref{tab:mean_z_binned}. The deviations in redshift between the tomographically-binned samples are smaller than in the unbinned case, but still statistically significant (except for the case of the $0.4<z<0.6$ bin, in which the one-to-one sample is consistent with the all observed sample). 

The DESC SRD specifies the maximum uncertainty on the mean tomographic redshift to be $\sigma_{\bar{z}}<0.005(1 + \bar{z})$. Our observed deviations between the one-to-one and all observed samples are about 10\,\% of the upper bound on $\sigma_{\bar{z}}$. Biased estimates of $N(z)$ due to an isolated calibration sample may therefore contribute several percent of the DESC SRD redshift error budget. Numerous other systematic effects are also expected to bias $N(z)$, such as photometric redshift uncertainties and survey non-uniformity. If these effects correlate with blending, the impact could increase even further, but such analysis is beyond the scope of this work. Our results demonstrate the importance of minimizing and/or mitigating the impact of blending in cosmological analyses of LSST.

\def\arraystretch{1.3}
\begin{table}
    \centering
    \caption{Mean and median redshift values of our samples. For the observed catalog, we use the true redshift of the best match in the truth catalog (as described in Section \ref{ssec:matching}). Uncertainties are negligible.}
    \begin{tabular}{ccc}
        \hline
        Sample & Mean $z$ & Median $z$\\
        \hline
        One-to-One & $0.6982$ &$0.7092$\\
        Multiple-to-One & $0.7092$ &$0.7220$\\
        All Observed & $0.7042$ &$0.7160$\\
        Truth & $0.7021$ &$0.7141$\\
        \hline
    \end{tabular}
    \label{tab:table1}
\end{table}

\begin{table*}
    \centering
    \caption{Mean redshift values for our tomographically-binned samples. For the observed catalog, we use the true redshift of the best match in the truth catalog (as described in Section \ref{ssec:matching}). Uncertainties are negligible. We also show the deviation of the one-to-one sample from the all observed sample.}
    \begin{tabular}{cccccc}
        \multicolumn{6}{c}{Mean Redshift}\\
        \hline
        Redshift & One-to-One & Multiple-to-One & All Observed & Truth & $z_{\textrm{all}}-z_{\textrm{1-to-1}}$\\
        \hline
        0.2-0.4 & $0.3078$ & $0.3079$ & $0.3080$ & $0.3080$ & $0.0002$\\
        0.4-0.6 & $0.5036$ & $0.5037$ & $0.5036$ & $0.5036$ & $0.0000$\\
        0.6-0.8 & $0.7037$ & $0.7039$ & $0.7038$ & $0.7037$ & $0.0001$\\
        0.8-1.0 & $0.8936$ & $0.8947$ & $0.8943$ & $0.8939$ & $0.0007$\\
        1.0-1.2 & $1.0890$ & $1.0908$ & $1.0901$ & $1.0897$ & $0.0011$\\
        \hline
    \end{tabular}
    
    \label{tab:mean_z_binned}
\end{table*}

\section{Bias in Galaxy Clustering}
\label{sec:w_results}
\subsection{Measuring the galaxy clustering signal}
\label{ssec:measurement}

In LSST, two-point measurements of galaxy clustering are planned to be a pillar of the so-called ``$3\times2$-point'' cosmology analyses~\citep{KiDS1000_cosmo, DESY3_cosmo, HSCY3}, along with galaxy-galaxy lensing (two-point correlation of galaxy position and galaxy weak lensing) and cosmic shear (two-point correlation of galaxy weak lensing).

In this work we use the Landy \& Szalay estimator \citep{1993ApJ...412...64L} to estimate the two-point correlation function of the galaxy density field:

\begin{equation}
    w(\theta) = \frac{DD(\theta) - 2DR(\theta) + RR(\theta)}{RR(\theta)},
\end{equation}
where $D$ stands for ``data'', or the galaxy sample that traces large-scale structure, and $R$ stands for ``random'', a sample of random points that occupy the same footprint as $D$ but are spatially randomly distributed. $DD(\theta)$ are the number of data-data pairs separated by an angle $\theta$, $DR(\theta)$ is the number of data-random pairs, and $RR(\theta)$ is the number of random-random pairs. The correlation functions are estimated using \textsc{TreeCorr}~\citep{2004MNRAS.352..338J} and are computed in 12 log-spaced angular bins between $0.1'$ and $250'$. For our nominal analysis we will focus on scales where $k < 0.3 h$\,Mpc$^{-1}$, following the DESC SRD.

The covariance matrices are estimated via jackknife resampling on the larger SkySim5000 simulation\footnote{For additional information on SkySim5000 we refer the reader to \url{github.com/LSSTDESC/gcr-catalogs}.}, which expands the footprint of the CosmoDC2 catalog to 5000\,deg$^{2}$. We use 150 jackknife patches. SkySim5000 is not an end-to-end image simulation, but it is a direct extension of the truth catalog; therefore, it allows us to obtain a covariance estimate that is as close as possible to the truth covariance. This estimate also has lower statistical noise thanks to SkySim5000's larger footprint. We rescale the SkySim5000 covariances by the area ratio between the simulations in order to obtain the final DC2 covariance, using the so-called $f_{\rm{sky}}$ approximation \citep{Knox1995}. Due to the conservative nature of the dataset in this analysis and the simplistic geometry of our footprint, the contributions to the covariance matrix in our analysis are dominated by the cosmic variance component, and therefore SkySim5000 provides a good approximation to the covariance matrix for the analysis.

\subsection{Bias in the correlation function}
\label{ssec:wtheta_difference}

\begin{figure*}
    \includegraphics[width=2.1\columnwidth]{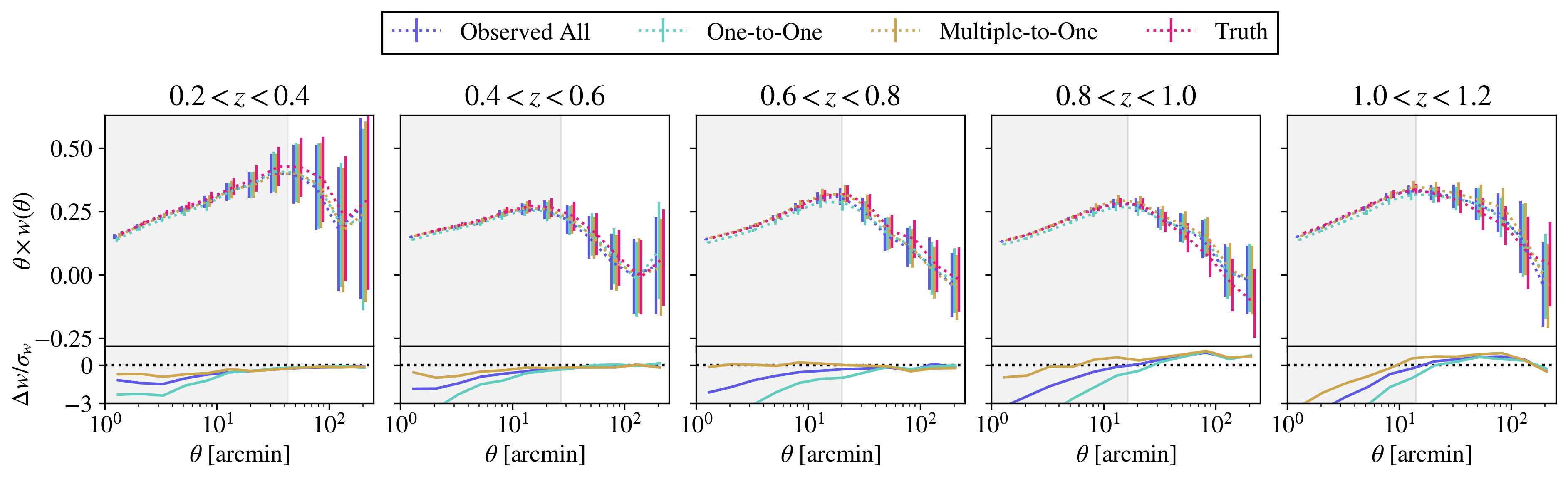}
    \caption{\emph{Top}: angular autocorrelation function measurements for each of our samples. \emph{Bottom}: deviation from the true correlation function. Error bars are computed as the square root of the covariance matrix diagonal, with the covariance matrix estimated from SkySim5000 as described in Section \ref{ssec:measurement}. Gray shaded regions denote small scales ignored in cosmological analyses, as specified by the DESC SRD ($k<0.3h$\,Mpc$^{-1}$). Note that the apparent redshift dependence on the bias in the bottom panel is largely dominated by the improved statistics in the higher redshift bins, which lead to reduced uncertainties.}
    \label{fig:angular_correlation}
\end{figure*}

We shift our focus to analyze and compare the two-point correlation functions obtained for the different galaxy samples (one-to-one, multiple-to-one, all observed, and the underlying truth galaxy catalog) that we introduce in Section~\ref{ssec:data}, across the five tomographic bins described in Section \ref{sec:z_results}. The results are shown in Figure \ref{fig:angular_correlation}. 

At small scales ($<10'$), all three observed samples exhibit less clustering than the truth; for the all observed sample (what we will measure in LSST) the deviation can be beyond $3\sigma$ as measured in DC2. This is the expected result of blending: the number count bias from blended objects that are ``lost'' from the sample will predominantly affect denser regions, lowering the average overall galaxy clustering signal. Note, however, that we cannot directly infer that blending is the cause of the deviations from the truth sample. Numerous observational effects (including, but not limited to, survey depth non-uniformity, background subtraction errors, and stellar density) may result in an overall bias between the truth and observed samples, although \cite{2020MNRAS.497..210S} has shown these impacts to be generally small. Another potential source of bias is the fact that the truth sample is not comprised of exactly the same objects as the observed sample; our truth catalog contains all galaxies with $i<24.1$ rather than only those that are matched with detections. All the aforementioned effects, however, should affect all three observed samples consistently. Therefore, the deviation from the truth sample may not be entirely caused by blending, but any deviation between the observed samples must be. Blending thus is responsible for deviations from the true correlation function at least up to the significance of the deviations between the observed samples.

Indeed, at small scales the multiple-to-one sample has a larger amplitude than the one-to-one sample. The correlation function of the all observed sample lies in between, as expected given that it is a superset of the two. These differences are expected given our definition of the multiple-to-one sample: projected alignments between objects will occur more often in environments with a higher clustering bias, resulting in a higher correlation function. The one-to-one sample, which selects isolated objects, should have a comparatively lower correlation function. The small-scale deviation between the one-to-one and multiple-to-one samples can be beyond $3\sigma$ as measured in DC2. We emphasize that, since our observed samples differ purely based on their blendedness, we can infer that the differences in their measured correlation functions are due to blending alone.

At larger scales, the observed samples are mostly consistent with the truth but show deviations on the order of $0.5\sigma$ in the higher two redshift bins. These deviations at large scales are consistent between the one-to-one and multiple-to-one samples, suggesting (as previously discussed) that the origin is likely not blending. A deviation below $1\sigma$ is acceptable here, and it is difficult—and irrelevant to our conclusions—to identify the exact observational effect that causes these deviations.

We extrapolate these DC2-based results to the full LSST area by rescaling the uncertainties by the area ratio, $A_{\rm{ratio}} = \sqrt{A_{\rm{LSST}}/A_{\rm{DC2}}} \approx 7.8$, where $A_{\rm{LSST}} = 18000$\,deg$^{2}$ and $A_{\rm{DC2}} = 294$\,deg$^{2}$. On small scales, this may translate into deviations of up to $21\sigma$ or larger. These deviations may complicate efforts to measure and model nonlinear galaxy bias~\citep{Nicola23} and other small scale effects \cite[e.g.,][]{2019OJAp....2E...4C}.

\subsection{Bias in inferred cosmology}
\label{ssec:cosmology}

We perform a cosmological analysis on our measured correlation functions following the analysis choices from the DESC SRD. We model the correlation function using the \textsc{Core Cosmology Library}~\citep[CCL,][]{2019ApJS..242....2C} with a linear galaxy bias for each redshift bin. First, we obtain the angular power spectrum:

\begin{equation}
C_{\ell} =  b^{2}(\bar{z})\frac{2}{2\ell +1}\int dz \left(\frac{dn}{dz}\right)^{2} H^{2}(z) P\left(k=\frac{\ell+1}{r(z)}, z\right),
\end{equation}
where $b(\bar{z})$ is the galaxy bias at the mean redshift of the sample, $dn/dz$ is the redshift distribution, $H(z)$ is the Hubble parameter, and $P(k)$ is the matter-power spectrum as computed by CAMB~\citep{Lewis:1999bs}. An inverse Fourier transform yields the theory correlation function:

\begin{equation}
w^{th}(\theta) = \sum_{\ell \geq 0}\left(\frac{2\ell+1}{4\pi}\right)\mathcal{P}_{\ell}(\cos{\theta})C_{\ell},
\end{equation}
where $\mathcal{P}_{\ell}$ is the Legendre polynomial of order $\ell$.

We obtain the posterior probability for the cosmological analysis by evaluating the Bayesian likelihood:

\begin{equation}\label{eqn:likelihood}
 \log{\mathcal{L}} = -0.5 \sum_{i}\sum_{j, k}\left(w_{i}(\theta_{j})-w^{th}_{i}(\theta_{j})\right) C^{-1}_{i, j k}\left(w_{i}(\theta_{k})-w_{i}^{th}(\theta_{k})\right), 
\end{equation}
where $i$ runs over the redshift bins and $j, k$ each run over the angular bins. $C^{-1}_{i, j k}$ is the inverse covariance matrix for the bin $i$. We ignore the correlations between redshift bins given that we are using top-hat bins in true redshift. However, we do not expect the conclusions from this work to be affected by these approximations. This is due to the fact that we are ignoring the effects in photometric redshifts, and that in a typical cosmological analyses the information comes from both the auto- and cross-power spectra; however, for the case of top-hat bins the same information is encoded in the auto spectra.

Using the package \textsc{emcee}~\citep{2013PASP..125..306F}, we run a Markov Chain Monte Carlo process (MCMC) to estimate the posterior distributions for $\Omega_{\rm m}$ and the galaxy bias parameters $b_{i}(\bar{z})$. We set truncated Gaussian priors for $\Omega_{\rm m}$ and $b_{i}$; the parameters of the Gaussian are as specified by the DESC SRD, while the truncation boundaries are set so as to prevent walkers from wandering beyond physically meaningful values. These priors are shown in Table \ref{tab:priors}. The free parameters in our model are those to which galaxy clustering is the most sensitive~\citep{Percival2001}. All other cosmological parameters are set to their DC2 simulation input values for the purposes of modeling the correlation function.

Our likelihood function is given in Equation \ref{eqn:likelihood}. For each chain we use gradient minimization to set starting values, run 32000 samples, and discard 100 samples per walker for burn-in. We confirm that the burn-in period is adequate and that the chains converge by visually inspecting the sample distributions to check that there is no evolution in mean and variance over time. To visualize the posteriors we use the Kernel Density Estimator function of \textsc{ChainConsumer} \citep{Hinton2016}.

\begin{table}
    \centering
    \caption{Top: priors used for our likelihood estimation. $\mathcal{U}(a, b)$ represents a uniform (top-hat) prior between $a$ and $b$. $\mathcal{N}(\mu, \sigma)$ represents a Gaussian prior with mean $\mu$ and standard deviation $\sigma$. Bottom: fixed values for cosmological parameters used in our model. The parameters are set to their DC2 input values \citep{2019ApJS..245...26K}.}
    \begin{tabular}{cc}
    \hline
    Parameter & Prior/Value\\
    \hline
    $\Omega_{\rm m}$ & $\mathcal{U}(0.1, 0.5) \times \mathcal{N}(0.2648, 0.2)$ \\
    $b_i$ & $\mathcal{U}(0.3, 4.0) \times \mathcal{N}(1.9, 0.9)$ \\
    \hline
    $\Omega_{\rm b}$ & $0.0448$ \\
    $\sigma_8$ & $0.8$ \\
    $h$ & $0.71$ \\
    $n_{\rm s}$ & $0.963$ \\
    \hline
    \end{tabular}
    \label{tab:priors}
\end{table}

\begin{figure*}
    \centering
    \includegraphics[width=\textwidth]{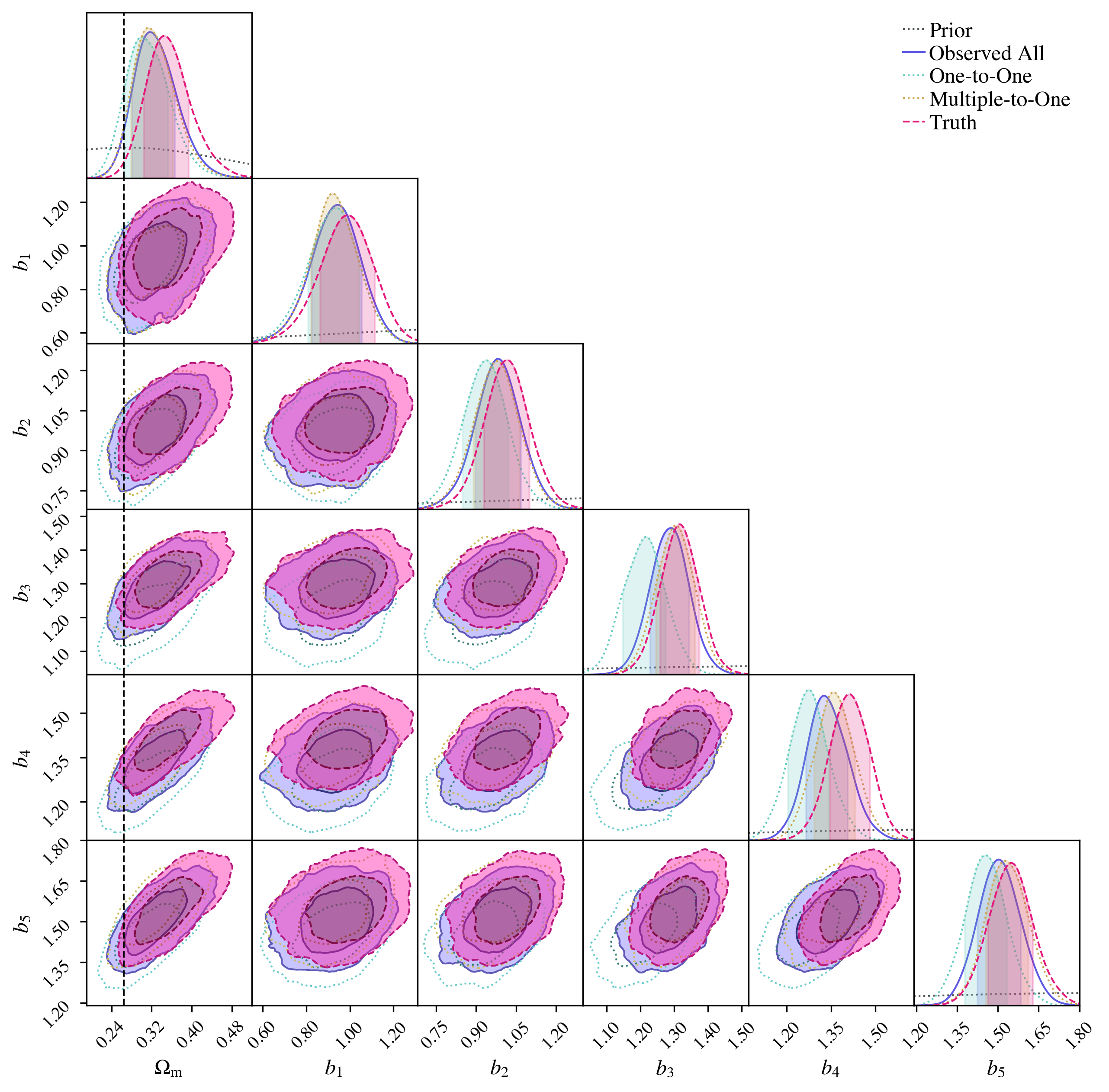}
    \caption{Contours for cosmological parameters assuming $\Lambda$CDM model with $\Omega_{\rm m}$ and galaxy bias $b_{i}$ freed, computed for each of our samples. We use DESC SRD priors and show the simulation input value of $\Omega_{\rm m}=0.265$ as a vertical dashed line.}
    \label{fig:5bin_posteriors}
\end{figure*}

The MCMC posteriors are shown in Figure \ref{fig:5bin_posteriors}. The cosmological constraints obtained for all of our observed samples are consistent with those obtained for the truth sample within $1\sigma$. Additionally, all of the constraints obtained for the galaxy samples considered here are consistent with the DC2 input ($\Omega_{\rm m}=0.265$) within  $ < 2\sigma$, although we find a small preference towards higher than expected $\Omega_{\rm m}$ values. In Appendix \ref{app:skysim} we  explore this deviation, finding that this preference is due to sample variance. The most significant differences in the inferred parameters between the samples are found in the galaxy biases, particularly in the one-to-one sample at high redshifts. This is expected, as the one-to-one sample is selected in a way that is naturally less clustered than the other two observed samples. We show the fractional uncertainty for the biases for each sample in Table \ref{tab:b-sig_b}. From the posterior distributions in Figure~\ref{fig:5bin_posteriors} we conclude that blending has no measurable impact on the inferred cosmology using the DC2 footprint with LSST Y1 fiducial analysis choices.

\begin{table}
    \centering
    \caption{Fractional deviation from the truth sample for each of the cosmological parameters ($\Delta X=X_{\rm{Truth}} - X_{\rm{Sample}}$).}
    \vspace{-1.3em}
    \begin{tabular}{cccc}
    \multicolumn{4}{c}{}\\
    \hline
    & One-to-One & Multiple-to-One & All Observed\\
    \hline
    $\Delta \Omega_{\rm m}/\sigma$ & 0.355 & 0.253 & 0.224\\
    $\Delta b_1/\sigma$ & 0.162 & 0.212 & 0.135\\
    $\Delta b_2/\sigma$ & 0.330 & 0.161 & 0.139\\
    $\Delta b_3/\sigma$ & 0.571 & 0.064 & 0.160\\
    $\Delta b_4/\sigma$ & 0.709 & 0.278 & 0.439\\
    $\Delta b_5/\sigma$ & 0.425 & 0.100 & 0.201\\
    \hline
    \end{tabular}

    \label{tab:b-sig_b}
\end{table}

\begin{figure}
    \centering
    \includegraphics[width=\columnwidth]{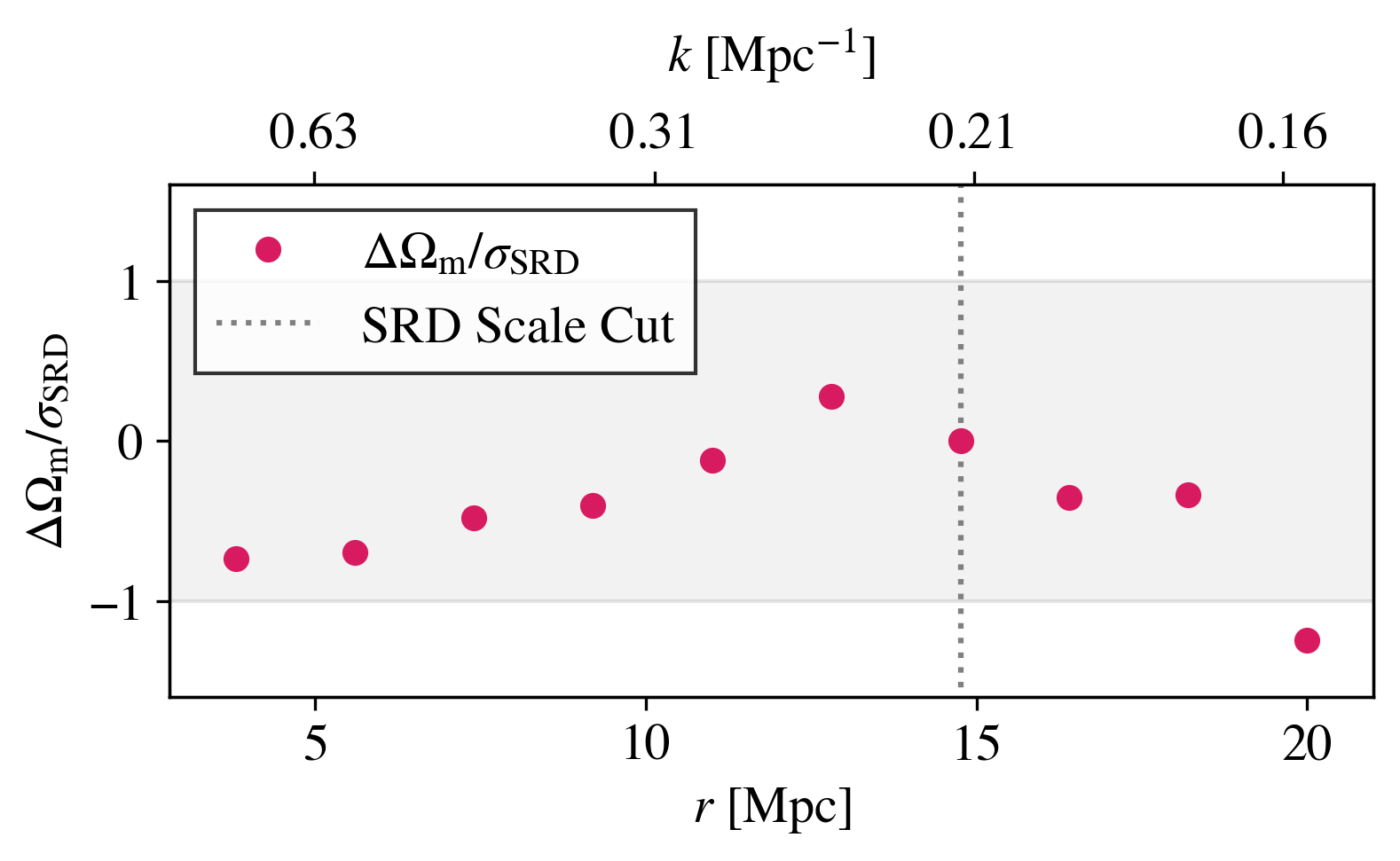}
    \caption{Relative deviation of best-fit $\Omega_{\rm m}$ from our fiducial cosmological analysis ($\Delta\Omega_{\rm m}=\Omega_{\rm m, \mathrm{best-fit}} - \Omega_{\rm m, \mathrm{SRD}}$) compared to the uncertainty at our fiducial scale cut ($\sigma_\mathrm{SRD}$), as a function of scale cut. The gray shaded region shows the 1$\sigma$ range. The DESC SRD scale cut used in our cosmological analysis is shown with a gray dotted line.}
    \label{fig:scale_cuts}
\end{figure}

As discussed in Section \ref{ssec:wtheta_difference}, the deviation between the truth and observed samples is amplified at smaller scales which are not used in the cosmological analysis. To test how the choice of small-scale cut could impact the inferred cosmology, we compute the best-fit $\Omega_{\rm m}$ for various scale cutoffs using gradient descent minimization. The results are shown in Figure \ref{fig:scale_cuts}; we find that the best-fit $\Omega_{\rm m}$ value at smaller scale cuts is compatible with our fiducial analysis within 1$\sigma$ for all but the most conservative scale cut (20\,Mpc). Note, however, that the increased uncertainty on the inferred cosmology for the 20\,Mpc scale cut would likely bring it into 1$\sigma$ concordance with our fiducial analysis as well.

\section{Limitations}
\label{sec:limitations}
Our fiducial analysis includes two major simplifications to the data set. First, we use the true redshifts of each observed galaxy, generated by matching the observed catalog to the true catalog. Photometric redshifts will add a further complication to the measurements. Second, we perform a magnitude cut at the Y1 Gold limit, despite the fact that DC2 is measured at a Y5 depth.

To test whether these simplifications affect our conclusions, we re-run the analysis using the LSST Y5 magnitude cut ($i<24.92$) and a BPZ photometric redshift catalog for DC2. Our main results hold in these secondary analyses. The procedure and results of these tests are described in detail in Appendix \ref{app:results_photz}.

These tests provide evidence that blending will not significantly impact $\Omega_{\rm m}$ measurements from galaxy clustering analyses in the context of a Y5 analysis with photometric redshift effects. However, for the reasons discussed in Section \ref{sec:intro}, it is difficult to isolate the exact impact of blending in these measurements as opposed to the effects of incompleteness and biases introduced by the photometric redshift algorithm itself.

In addition to these simplifications, we also highlight three additional areas for future study. First, most cosmological analyses performed to date limit the usage of galaxy-clustering results to linear scales~\citep[e.g.,][]{KiDS1000_cosmo, DESY3_cosmo} on their baseline analyses. However, there are significant efforts in the community to improve our understanding of small scales~\citep{Nicola23} and unlock additional cosmological information.
In our baseline analysis we focus on linear scales and the impact of blending on the linear bias. However, Figure~\ref{fig:angular_correlation} clearly shows that the impact of blending is more significant at small (non-linear) scales. Thus, beyond improvements in modeling non-linearities, additional work understanding and modeling the impact of blending at these scales will be required in order to push further into the non-linear regime using LSST data.

Second, we consider galaxy clustering in isolation and do not address the effect of blending in the full $3\times 2$pt analyses. \citet{Nourbakhsh2022} find significant impact in cosmic-shear-only analyses, and our results point to significant impact in clustering-only analyses for the full LSST area (though not in the smaller footprint of DC2), which motivates a deeper study of the impact in $3\times 2$pt analyses.

Finally, we note that the galaxies used for the clustering analysis are detected and deblended using a particular detection and deblending algorithm (those included in the Rubin Science Pipelines v19.0.0), and the specifics of the effect of blending found in the two-point correlation function do depend ultimately on these. However, potential improvements in these algorithms and/or combinations with space-based higher-resolution imaging can mitigate the impact of blending in galaxy clustering measurements found in our work.

\section{Summary \& Conclusions}
\label{sec:conclusions}
In this work we study the impact of the number count bias from blending in two-point galaxy clustering analyses for the Vera C. Rubin Observatory LSST.\footnote{The code used for our analysis of catalogs, correlation functions, and cosmology, as well as plotting routines, is available at \url{github.com/bclevine/BLxClustering}.}

Using data from the DC2 simulated sky survey gives us the opportunity to analyze LSST-like data while still having access to the ground truth, which provides an ideal scenario to test the effects of blending. After comparing and matching with the truth catalog we subdivide the observed galaxies into two main samples: the one-to-one matches, i.e., observed galaxies that have a single neighbor in the truth catalog within a $R_{\rm{max}}=1''$ radius; and the multiple-to-one matches, i.e., those observed galaxies that have more than one neighbor in the truth catalog within a $R_{\rm{max}}=1''$ radius. These two categories roughly correspond to likely-not-blended and likely-blended galaxies, respectively.  By using a distance-based classification criterion we attempt to make our conclusions as deblender-independent as possible.

We follow the fiducial Y1 analysis presented in the DESC SRD and analyze potential differences between the one-to-one sample, the multiple-to-one sample, and the combination of both (the all observed sample), as well as the underlying truth. We first look for differences in the ensemble redshift distributions, finding that the one-to-one sample has a different mean redshift than the multiple-to-one and the all observed samples. These biases, while statistically significant, will likely be subdominant compared to other sources of mean redshift bias such as photometric redshift errors. Further study of how different redshift calibration algorithms interact with blending will be required in order to quantify the full impact to $N(z)$ measurements.

Second, we compare the two-point correlation functions obtained for the one-to-one, multiple-to-one, all observed, and truth galaxies. We find that at small, non-linear scales the differences between the observed and truth correlation functions are highly significant $(> 3 \sigma)$, but at linear scales we do not find any measurable systematic biases due to blending. When extrapolating these results to the full LSST area, these differences at small scales become extremely significant ($> 21\sigma$). These differences advise for caution when pushing to the small-scale regime using galaxy clustering two-point measurements in future cosmological analyses, as systematic biases may arise if the effect of blending is not accounted for properly. We find some mild deviation in the measured correlation functions between the observed galaxies and the truth at linear scales. However, these are most likely not due to blending, but rather to other observational effects, as the three observed samples (one-to-one, multiple-to-one, and all observed) all exhibit the same effect.

We use the measured two-point data vectors to obtain constraints on the combination of $\Omega_{\rm m}$ and galaxy bias parameters. We find that the inferred posterior distributions for $\Omega_{\rm m}$ for all samples (one-to-one, multiple-to-one, all observed, and truth) are statistically compatible. Furthermore, we obtain best-fit $\Omega_{\rm m}$ values statistically compatible with the input $\Omega_{\rm m}=0.265$ used for DC2. The inferred best-fit $\Omega_{\rm m}$ values are slightly higher than the simulation input, but we confirm that this is due to sample variance present in the DC2 area. We obtain the same results from cosmology analyses using a Y5 magnitude cut and photometric redshifts.

Our results provide evidence that, for the LSST Y1 Gold sample with the fiducial scale cuts, the number count bias from blending will not significantly impact the results of galaxy clustering analyses. However, for cosmology beyond the fiducial Y1 analysis, it will be important to 1) measure the correlation of blending with other sources of redshift bias to confirm that blending remains subdominant; and 2) study and model the impact of blending on angular scales smaller than our cutoff of $k=0.3h$\,Mpc$^{-1}$, allowing us to take advantage of the statistical power of the correlation function on nonlinear scales.

\section*{Acknowledgments}

This paper has undergone internal review in the LSST Dark Energy Science Collaboration. The internal reviewers were Cyrille Doux, Eric Gawiser, and Boris Leistedt. The authors acknowledge the usage of \textsc{Astropy, CCL, emcee, healpy, Matplotlib, NumPy, pandas, scikit-learn, SciPy, SkyProj,} and \textsc{TreeCorr}. B. Levine and A. von der Linden are supported by DOE grants DE-SC0023387 and DE-SC0025309. C. Chang is supported by DOE grant DE-SC0021949. E. Gawiser is supported by DOE grant DE-SC0010008. B. Leistedt is supported by the Royal Society through a University Research Fellowship.

The DESC acknowledges ongoing support from the Institut National de Physique Nucl\'eaire et de Physique des Particules in France; the Science \& Technology Facilities Council in the United Kingdom; and the Department of Energy and the LSST Discovery Alliance in the United States. DESC uses resources of the IN2P3 Computing Center (CC-IN2P3--Lyon/Villeurbanne - France) funded by the Centre National de la Recherche Scientifique; the National Energy Research Scientific Computing Center, a DOE Office of Science User Facility supported by the Office of Science of the U.S.\ Department of Energy under Contract No.\ DE-AC02-05CH11231; STFC DiRAC HPC Facilities, funded by UK BEIS National E-infrastructure capital grants; and the UK particle physics grid, supported by the GridPP Collaboration.  This work was performed in part under DOE Contract DE-AC02-76SF00515. This manuscript has been authored by Fermi Research Alliance, LLC under Contract No. DE-AC02-07CH11359 with the U.S. Department of Energy, Office of Science, Office of High Energy Physics.

The contributions from the authors are as follows. B. Levine performed the main analysis of redshift, angular correlation function, and cosmological measurements. J. S\'{a}nchez supervised the project, performed the nearest-neighbors matching, and produced correlation function measurements for SkySim5000 and some instances of DC2. B. Levine and J. S\'{a}nchez led the paper writing. J. S\'{a}nchez and C. Chang proposed the project and organized collaborators. C. Chang and A. von der Linden provided guidance on both the analysis and text. E. Gawiser and B. Leistedt offered extensive feedback as internal reviewers, leading to paper updates. C. Chang, E. Gawiser, and J. S\'{a}nchez are DESC builders. K. Krzy\.za\'{n}ska and E. Collins contributed to initial stages of the project and initial analyses.

\bibliographystyle{mnras}

\bibliography{blending}

\begin{thebibliography}{}
\makeatletter
\relax
\def\mn@urlcharsother{\let\do\@makeother \do\$\do\&\do\#\do\^\do\_\do\%\do\~}
\def\mn@doi{\begingroup\mn@urlcharsother \@ifnextchar [ {\mn@doi@} {\mn@doi@[]}}
\def\mn@doi@[#1]#2{\def\@tempa{#1}\ifx\@tempa\@empty \href {http://dx.doi.org/#2} {doi:#2}\else \href {http://dx.doi.org/#2} {#1}\fi \endgroup}
\def\mn@eprint#1#2{\mn@eprint@#1:#2::\@nil}
\def\mn@eprint@arXiv#1{\href {http://arxiv.org/abs/#1} {{\tt arXiv:#1}}}
\def\mn@eprint@dblp#1{\href {http://dblp.uni-trier.de/rec/bibtex/#1.xml} {dblp:#1}}
\def\mn@eprint@#1:#2:#3:#4\@nil{\def\@tempa {#1}\def\@tempb {#2}\def\@tempc {#3}\ifx \@tempc \@empty \let \@tempc \@tempb \let \@tempb \@tempa \fi \ifx \@tempb \@empty \def\@tempb {arXiv}\fi \@ifundefined {mn@eprint@\@tempb}{\@tempb:\@tempc}{\expandafter \expandafter \csname mn@eprint@\@tempb\endcsname \expandafter{\@tempc}}}

\bibitem[\protect\citeauthoryear{{Abbott} et~al.,}{{Abbott} et~al.}{2018}]{2018PhRvD..98d3526A}
{Abbott} T.~M.~C.,  et~al., 2018, \mn@doi [\prd] {10.1103/PhysRevD.98.043526}, \href {https://ui.adsabs.harvard.edu/abs/2018PhRvD..98d3526A} {98, 043526}

\bibitem[\protect\citeauthoryear{{Abbott} et~al.,}{{Abbott} et~al.}{2022}]{DESY3_cosmo}
{Abbott} T.~M.~C.,  et~al., 2022, \mn@doi [\prd] {10.1103/PhysRevD.105.023520}, \href {https://ui.adsabs.harvard.edu/abs/2022PhRvD.105b3520A} {105, 023520}

\bibitem[\protect\citeauthoryear{{Aihara} et~al.,}{{Aihara} et~al.}{2018}]{Aihara2017}
{Aihara} H.,  et~al., 2018, \mn@doi [\pasj] {10.1093/pasj/psx081}, \href {https://ui.adsabs.harvard.edu/abs/2018PASJ...70S...8A} {70, S8}

\bibitem[\protect\citeauthoryear{{Akeson} et~al.,}{{Akeson} et~al.}{2019}]{Roman2019}
{Akeson} R.,  et~al., 2019, \mn@doi [arXiv e-prints] {10.48550/arXiv.1902.05569}, \href {https://ui.adsabs.harvard.edu/abs/2019arXiv190205569A} {p. arXiv:1902.05569}

\bibitem[\protect\citeauthoryear{{Albrecht}}{{Albrecht}}{2006}]{2006astro.ph..9591A}
{Albrecht} A.,  2006, in APS April Meeting Abstracts. APS Meeting Abstracts.
p. G1.002

\bibitem[\protect\citeauthoryear{{Arcelin}, {Doux}, {Aubourg}, {Roucelle}  \& {LSST Dark Energy Science Collaboration}}{{Arcelin} et~al.}{2021}]{2021MNRAS.500..531A}
{Arcelin} B.,  {Doux} C.,  {Aubourg} E.,  {Roucelle} C.,   {LSST Dark Energy Science Collaboration} 2021, \mn@doi [\mnras] {10.1093/mnras/staa3062}, \href {https://ui.adsabs.harvard.edu/abs/2021MNRAS.500..531A} {500, 531}

\bibitem[\protect\citeauthoryear{{Behroozi}, {Wechsler}, {Hearin}  \& {Conroy}}{{Behroozi} et~al.}{2019}]{2019MNRAS.488.3143B}
{Behroozi} P.,  {Wechsler} R.~H.,  {Hearin} A.~P.,   {Conroy} C.,  2019, \mn@doi [\mnras] {10.1093/mnras/stz1182}, \href {https://ui.adsabs.harvard.edu/abs/2019MNRAS.488.3143B} {488, 3143}

\bibitem[\protect\citeauthoryear{{Ben{\'\i}tez}}{{Ben{\'\i}tez}}{2000}]{BPZ2000}
{Ben{\'\i}tez} N.,  2000, \mn@doi [\apj] {10.1086/308947}, \href {https://ui.adsabs.harvard.edu/abs/2000ApJ...536..571B} {536, 571}

\bibitem[\protect\citeauthoryear{{Benson}}{{Benson}}{2012}]{2012NewA...17..175B}
{Benson} A.~J.,  2012, \mn@doi [\na] {10.1016/j.newast.2011.07.004}, \href {https://ui.adsabs.harvard.edu/abs/2012NewA...17..175B} {17, 175}

\bibitem[\protect\citeauthoryear{{Bernstein}, {Armstrong}, {Krawiec}  \& {March}}{{Bernstein} et~al.}{2016}]{2016MNRAS.459.4467B}
{Bernstein} G.~M.,  {Armstrong} R.,  {Krawiec} C.,   {March} M.~C.,  2016, \mn@doi [\mnras] {10.1093/mnras/stw879}, \href {https://ui.adsabs.harvard.edu/abs/2016MNRAS.459.4467B} {459, 4467}

\bibitem[\protect\citeauthoryear{{Bosch} et~al.,}{{Bosch} et~al.}{2018}]{Bosch2018}
{Bosch} J.,  et~al., 2018, \mn@doi [\pasj] {10.1093/pasj/psx080}, \href {https://ui.adsabs.harvard.edu/abs/2018PASJ...70S...5B} {70, S5}

\bibitem[\protect\citeauthoryear{{Cawthon} et~al.,}{{Cawthon} et~al.}{2022}]{Cawthon2022}
{Cawthon} R.,  et~al., 2022, \mn@doi [\mnras] {10.1093/mnras/stac1160}, \href {https://ui.adsabs.harvard.edu/abs/2022MNRAS.513.5517C} {513, 5517}

\bibitem[\protect\citeauthoryear{{Chisari} et~al.,}{{Chisari} et~al.}{2019a}]{2019OJAp....2E...4C}
{Chisari} N.~E.,  et~al., 2019a, \mn@doi [The Open Journal of Astrophysics] {10.21105/astro.1905.06082}, \href {https://ui.adsabs.harvard.edu/abs/2019OJAp....2E...4C} {2, 4}

\bibitem[\protect\citeauthoryear{{Chisari} et~al.,}{{Chisari} et~al.}{2019b}]{2019ApJS..242....2C}
{Chisari} N.~E.,  et~al., 2019b, \mn@doi [\apjs] {10.3847/1538-4365/ab1658}, \href {https://ui.adsabs.harvard.edu/abs/2019ApJS..242....2C} {242, 2}

\bibitem[\protect\citeauthoryear{{Dawson}, {Schneider}, {Tyson}  \& {Jee}}{{Dawson} et~al.}{2016}]{2016ApJ...816...11D}
{Dawson} W.~A.,  {Schneider} M.~D.,  {Tyson} J.~A.,   {Jee} M.~J.,  2016, \mn@doi [\apj] {10.3847/0004-637X/816/1/11}, \href {https://ui.adsabs.harvard.edu/abs/2016ApJ...816...11D} {816, 11}

\bibitem[\protect\citeauthoryear{Du}{Du}{2023}]{Du_2023}
Du Z.,  2023, PhD thesis, UC Riverside

\bibitem[\protect\citeauthoryear{{Flaugher}}{{Flaugher}}{2005}]{Flaugher2005}
{Flaugher} B.,  2005, \mn@doi [International Journal of Modern Physics A] {10.1142/S0217751X05025917}, \href {http://adsabs.harvard.edu/abs/2005IJMPA..20.3121F} {20, 3121}

\bibitem[\protect\citeauthoryear{{Foreman-Mackey}, {Hogg}, {Lang}  \& {Goodman}}{{Foreman-Mackey} et~al.}{2013}]{2013PASP..125..306F}
{Foreman-Mackey} D.,  {Hogg} D.~W.,  {Lang} D.,   {Goodman} J.,  2013, \mn@doi [\pasp] {10.1086/670067}, \href {https://ui.adsabs.harvard.edu/abs/2013PASP..125..306F} {125, 306}

\bibitem[\protect\citeauthoryear{{Garc{\'\i}a-Garc{\'\i}a}, {Alonso}, {Ferreira}, {Hadzhiyska}, {Nicola}, {S{\'a}nchez}  \& {Slosar}}{{Garc{\'\i}a-Garc{\'\i}a} et~al.}{2023}]{GarciaGarcia2023}
{Garc{\'\i}a-Garc{\'\i}a} C.,  {Alonso} D.,  {Ferreira} P.~G.,  {Hadzhiyska} B.,  {Nicola} A.,  {S{\'a}nchez} C.,   {Slosar} A.,  2023, \mn@doi [\jcap] {10.1088/1475-7516/2023/01/025}, \href {https://ui.adsabs.harvard.edu/abs/2023JCAP...01..025G} {2023, 025}

\bibitem[\protect\citeauthoryear{{Gawiser} et~al.,}{{Gawiser} et~al.}{2006}]{Gawiser2006}
{Gawiser} E.,  et~al., 2006, \mn@doi [\apjs] {10.1086/497644}, \href {https://ui.adsabs.harvard.edu/abs/2006ApJS..162....1G} {162, 1}

\bibitem[\protect\citeauthoryear{{Gruen} et~al.,}{{Gruen} et~al.}{2019}]{2019MNRAS.488.4389G}
{Gruen} D.,  et~al., 2019, \mn@doi [\mnras] {10.1093/mnras/stz2036}, \href {https://ui.adsabs.harvard.edu/abs/2019MNRAS.488.4389G} {488, 4389}

\bibitem[\protect\citeauthoryear{{Heitmann} et~al.,}{{Heitmann} et~al.}{2019}]{2019ApJS..245...16H}
{Heitmann} K.,  et~al., 2019, \mn@doi [\apjs] {10.3847/1538-4365/ab4da1}, \href {https://ui.adsabs.harvard.edu/abs/2019ApJS..245...16H} {245, 16}

\bibitem[\protect\citeauthoryear{{Heymans} et~al.,}{{Heymans} et~al.}{2021}]{KiDS1000_cosmo}
{Heymans} C.,  et~al., 2021, \mn@doi [\aap] {10.1051/0004-6361/202039063}, \href {https://ui.adsabs.harvard.edu/abs/2021A&A...646A.140H} {646, A140}

\bibitem[\protect\citeauthoryear{{Hinton}}{{Hinton}}{2016}]{Hinton2016}
{Hinton} S.~R.,  2016, \mn@doi [The Journal of Open Source Software] {10.21105/joss.00045}, \href {http://adsabs.harvard.edu/abs/2016JOSS....1...45H} {1, 00045}

\bibitem[\protect\citeauthoryear{{Ivezi{\'c}} et~al.,}{{Ivezi{\'c}} et~al.}{2019}]{2019ApJ...873..111I}
{Ivezi{\'c}} {\v{Z}}.,  et~al., 2019, \mn@doi [\apj] {10.3847/1538-4357/ab042c}, \href {https://ui.adsabs.harvard.edu/abs/2019ApJ...873..111I} {873, 111}

\bibitem[\protect\citeauthoryear{{Jarvis}, {Bernstein}  \& {Jain}}{{Jarvis} et~al.}{2004}]{2004MNRAS.352..338J}
{Jarvis} M.,  {Bernstein} G.,   {Jain} B.,  2004, \mn@doi [\mnras] {10.1111/j.1365-2966.2004.07926.x}, \href {https://ui.adsabs.harvard.edu/abs/2004MNRAS.352..338J} {352, 338}

\bibitem[\protect\citeauthoryear{{Jones} \& {Heavens}}{{Jones} \& {Heavens}}{2019}]{2019MNRAS.483.2487J}
{Jones} D.~M.,  {Heavens} A.~F.,  2019, \mn@doi [\mnras] {10.1093/mnras/sty3279}, \href {https://ui.adsabs.harvard.edu/abs/2019MNRAS.483.2487J} {483, 2487}

\bibitem[\protect\citeauthoryear{{Joseph}, {Courbin}  \& {Starck}}{{Joseph} et~al.}{2016}]{2016A&A...589A...2J}
{Joseph} R.,  {Courbin} F.,   {Starck} J.~L.,  2016, \mn@doi [\aap] {10.1051/0004-6361/201527923}, \href {https://ui.adsabs.harvard.edu/abs/2016A&A...589A...2J} {589, A2}

\bibitem[\protect\citeauthoryear{{Juri{\'c}} et~al.,}{{Juri{\'c}} et~al.}{2017}]{RubinPipelines2015}
{Juri{\'c}} M.,  et~al., 2017, in Astronomical Data Analysis Software and Systems XXV. p.~279 (\mn@eprint {arXiv} {1512.07914}), \mn@doi{10.48550/arXiv.1512.07914}

\bibitem[\protect\citeauthoryear{{Knox}}{{Knox}}{1995}]{Knox1995}
{Knox} L.,  1995, \mn@doi [\prd] {10.1103/PhysRevD.52.4307}, \href {https://ui.adsabs.harvard.edu/abs/1995PhRvD..52.4307K} {52, 4307}

\bibitem[\protect\citeauthoryear{{Komatsu} et~al.,}{{Komatsu} et~al.}{2011}]{2011ApJS..192...18K}
{Komatsu} E.,  et~al., 2011, \mn@doi [\apjs] {10.1088/0067-0049/192/2/18}, \href {https://ui.adsabs.harvard.edu/abs/2011ApJS..192...18K} {192, 18}

\bibitem[\protect\citeauthoryear{{Korytov} et~al.,}{{Korytov} et~al.}{2019}]{2019ApJS..245...26K}
{Korytov} D.,  et~al., 2019, \mn@doi [\apjs] {10.3847/1538-4365/ab510c}, \href {https://ui.adsabs.harvard.edu/abs/2019ApJS..245...26K} {245, 26}

\bibitem[\protect\citeauthoryear{{Kovacs} et~al.,}{{Kovacs} et~al.}{2022}]{Kovacs2022}
{Kovacs} E.,  et~al., 2022, \mn@doi [The Open Journal of Astrophysics] {10.21105/astro.2110.03769}, \href {https://ui.adsabs.harvard.edu/abs/2022OJAp....5E...1K} {5, 1}

\bibitem[\protect\citeauthoryear{{LSST Dark Energy Science Collaboration} et~al.,}{{LSST Dark Energy Science Collaboration} et~al.}{2021}]{2020arXiv201005926L}
{LSST Dark Energy Science Collaboration} et~al., 2021, \mn@doi [\apjs] {10.3847/1538-4365/abd62c}, \href {https://ui.adsabs.harvard.edu/abs/2021ApJS..253...31L} {253, 31}

\bibitem[\protect\citeauthoryear{{Landy} \& {Szalay}}{{Landy} \& {Szalay}}{1993}]{1993ApJ...412...64L}
{Landy} S.~D.,  {Szalay} A.~S.,  1993, \mn@doi [\apj] {10.1086/172900}, \href {https://ui.adsabs.harvard.edu/abs/1993ApJ...412...64L} {412, 64}

\bibitem[\protect\citeauthoryear{{Laureijs} et~al.,}{{Laureijs} et~al.}{2011}]{Euclid2011}
{Laureijs} R.,  et~al., 2011, \mn@doi [arXiv e-prints] {10.48550/arXiv.1110.3193}, \href {https://ui.adsabs.harvard.edu/abs/2011arXiv1110.3193L} {p. arXiv:1110.3193}

\bibitem[\protect\citeauthoryear{Lewis, Challinor  \& Lasenby}{Lewis et~al.}{2000}]{Lewis:1999bs}
Lewis A.,  Challinor A.,   Lasenby A.,  2000, \mn@doi [\apj] {10.1086/309179}, 538, 473

\bibitem[\protect\citeauthoryear{{Lima}, {Cunha}, {Oyaizu}, {Frieman}, {Lin}  \& {Sheldon}}{{Lima} et~al.}{2008}]{2008MNRAS.390..118L}
{Lima} M.,  {Cunha} C.~E.,  {Oyaizu} H.,  {Frieman} J.,  {Lin} H.,   {Sheldon} E.~S.,  2008, \mn@doi [\mnras] {10.1111/j.1365-2966.2008.13510.x}, \href {https://ui.adsabs.harvard.edu/abs/2008MNRAS.390..118L} {390, 118}

\bibitem[\protect\citeauthoryear{MacCrann et~al.,}{MacCrann et~al.}{2021}]{MacCrann_2021}
MacCrann N.,  et~al., 2021, \mn@doi [Monthly Notices of the Royal Astronomical Society] {10.1093/mnras/stab2870}, 509, 3371–3394

\bibitem[\protect\citeauthoryear{Mandelbaum}{Mandelbaum}{2018}]{Mandelbaum_2018}
Mandelbaum R.,  2018, \mn@doi [Annual Review of Astronomy and Astrophysics] {10.1146/annurev-astro-081817-051928}, 56, 393

\bibitem[\protect\citeauthoryear{{Melchior}, {Moolekamp}, {Jerdee}, {Armstrong}, {Sun}, {Bosch}  \& {Lupton}}{{Melchior} et~al.}{2018}]{2018A&C....24..129M}
{Melchior} P.,  {Moolekamp} F.,  {Jerdee} M.,  {Armstrong} R.,  {Sun} A.~L.,  {Bosch} J.,   {Lupton} R.,  2018, \mn@doi [Astronomy and Computing] {10.1016/j.ascom.2018.07.001}, \href {https://ui.adsabs.harvard.edu/abs/2018A&C....24..129M} {24, 129}

\bibitem[\protect\citeauthoryear{{Melchior}, {Joseph}, {Sanchez}, {MacCrann}  \& {Gruen}}{{Melchior} et~al.}{2021}]{Melchior2021}
{Melchior} P.,  {Joseph} R.,  {Sanchez} J.,  {MacCrann} N.,   {Gruen} D.,  2021, \mn@doi [Nature Reviews Physics] {10.1038/s42254-021-00353-y}, \href {https://ui.adsabs.harvard.edu/abs/2021NatRP...3..712M} {3, 712}

\bibitem[\protect\citeauthoryear{{Merlin} et~al.,}{{Merlin} et~al.}{2015}]{2015A&A...582A..15M}
{Merlin} E.,  et~al., 2015, \mn@doi [\aap] {10.1051/0004-6361/201526471}, \href {https://ui.adsabs.harvard.edu/abs/2015A&A...582A..15M} {582, A15}

\bibitem[\protect\citeauthoryear{{Nicola} et~al.,}{{Nicola} et~al.}{2024}]{Nicola23}
{Nicola} A.,  et~al., 2024, \mn@doi [\jcap] {10.1088/1475-7516/2024/02/015}, \href {https://ui.adsabs.harvard.edu/abs/2024JCAP...02..015N} {2024, 015}

\bibitem[\protect\citeauthoryear{{Nourbakhsh}, {Tyson}, {Schmidt}  \& {LSST Dark Energy Science Collaboration}}{{Nourbakhsh} et~al.}{2022}]{Nourbakhsh2022}
{Nourbakhsh} E.,  {Tyson} J.~A.,  {Schmidt} S.~J.,   {LSST Dark Energy Science Collaboration} 2022, \mn@doi [\mnras] {10.1093/mnras/stac1303}, \href {https://ui.adsabs.harvard.edu/abs/2022MNRAS.514.5905N} {514, 5905}

\bibitem[\protect\citeauthoryear{{Pedregosa} et~al.,}{{Pedregosa} et~al.}{2011}]{scikit-learn}
{Pedregosa} F.,  et~al., 2011, \mn@doi [Journal of Machine Learning Research] {10.48550/arXiv.1201.0490}, \href {https://ui.adsabs.harvard.edu/abs/2011JMLR...12.2825P} {12, 2825}

\bibitem[\protect\citeauthoryear{{Percival} et~al.,}{{Percival} et~al.}{2001}]{Percival2001}
{Percival} W.~J.,  et~al., 2001, \mn@doi [\mnras] {10.1046/j.1365-8711.2001.04827.x}, \href {https://ui.adsabs.harvard.edu/abs/2001MNRAS.327.1297P} {327, 1297}

\bibitem[\protect\citeauthoryear{{Ramel}, {Doux}  \& {Kuna}}{{Ramel} et~al.}{2023}]{Ramel:2023fgr}
{Ramel} M.,  {Doux} C.,   {Kuna} M.,  2023. p. arXiv:2310.02079 (\mn@eprint {arXiv} {2310.02079}), \mn@doi{10.48550/arXiv.2310.02079}

\bibitem[\protect\citeauthoryear{{Ruiz-Zapatero}, {Hadzhiyska}, {Alonso}, {Ferreira}, {Garc{\'\i}a-Garc{\'\i}a}  \& {Mootoovaloo}}{{Ruiz-Zapatero} et~al.}{2023}]{RuizZapatero2023}
{Ruiz-Zapatero} J.,  {Hadzhiyska} B.,  {Alonso} D.,  {Ferreira} P.~G.,  {Garc{\'\i}a-Garc{\'\i}a} C.,   {Mootoovaloo} A.,  2023, \mn@doi [\mnras] {10.1093/mnras/stad1192}, \href {https://ui.adsabs.harvard.edu/abs/2023MNRAS.522.5037R} {522, 5037}

\bibitem[\protect\citeauthoryear{{S{\'a}nchez} et~al.,}{{S{\'a}nchez} et~al.}{2020}]{2020MNRAS.497..210S}
{S{\'a}nchez} J.,  et~al., 2020, \mn@doi [\mnras] {10.1093/mnras/staa1957}, \href {https://ui.adsabs.harvard.edu/abs/2020MNRAS.497..210S} {497, 210}

\bibitem[\protect\citeauthoryear{{Sanchez}, {Mendoza}, {Kirkby}, {Burchat}  \& {LSST Dark Energy Science Collaboration}}{{Sanchez} et~al.}{2021}]{2021arXiv210302078S}
{Sanchez} J.,  {Mendoza} I.,  {Kirkby} D.~P.,  {Burchat} P.~R.,   {LSST Dark Energy Science Collaboration} 2021, \mn@doi [\jcap] {10.1088/1475-7516/2021/07/043}, \href {https://ui.adsabs.harvard.edu/abs/2021JCAP...07..043S} {2021, 043}

\bibitem[\protect\citeauthoryear{{Schmidt} et~al.,}{{Schmidt} et~al.}{2020}]{2020MNRAS.499.1587S}
{Schmidt} S.~J.,  et~al., 2020, \mn@doi [\mnras] {10.1093/mnras/staa2799}, \href {https://ui.adsabs.harvard.edu/abs/2020MNRAS.499.1587S} {499, 1587}

\bibitem[\protect\citeauthoryear{{Sheldon} \& {Huff}}{{Sheldon} \& {Huff}}{2017}]{2017ApJ...841...24S}
{Sheldon} E.~S.,  {Huff} E.~M.,  2017, \mn@doi [\apj] {10.3847/1538-4357/aa704b}, \href {https://ui.adsabs.harvard.edu/abs/2017ApJ...841...24S} {841, 24}

\bibitem[\protect\citeauthoryear{{Sheldon}, {Becker}, {MacCrann}  \& {Jarvis}}{{Sheldon} et~al.}{2020}]{2020ApJ...902..138S}
{Sheldon} E.~S.,  {Becker} M.~R.,  {MacCrann} N.,   {Jarvis} M.,  2020, \mn@doi [\apj] {10.3847/1538-4357/abb595}, \href {https://ui.adsabs.harvard.edu/abs/2020ApJ...902..138S} {902, 138}

\bibitem[\protect\citeauthoryear{{Spergel} et~al.,}{{Spergel} et~al.}{2015}]{Roman2015}
{Spergel} D.,  et~al., 2015, \mn@doi [arXiv e-prints] {10.48550/arXiv.1503.03757}, \href {https://ui.adsabs.harvard.edu/abs/2015arXiv150303757S} {p. arXiv:1503.03757}

\bibitem[\protect\citeauthoryear{{Sugiyama} et~al.,}{{Sugiyama} et~al.}{2023}]{HSCY3}
{Sugiyama} S.,  et~al., 2023, \mn@doi [\prd] {10.1103/PhysRevD.108.123521}, \href {https://ui.adsabs.harvard.edu/abs/2023PhRvD.108l3521S} {108, 123521}

\bibitem[\protect\citeauthoryear{{The LSST Dark Energy Science Collaboration} et~al.,}{{The LSST Dark Energy Science Collaboration} et~al.}{2018}]{2018arXiv180901669T}
{The LSST Dark Energy Science Collaboration} et~al., 2018, arXiv e-prints, \href {https://ui.adsabs.harvard.edu/abs/2018arXiv180901669T} {p. arXiv:1809.01669}

\bibitem[\protect\citeauthoryear{Troxel et~al.,}{Troxel et~al.}{2023}]{Troxel_2023}
Troxel M.~A.,  et~al., 2023, \mn@doi [Monthly Notices of the Royal Astronomical Society] {10.1093/mnras/stad664}, 522, 2801–2820

\bibitem[\protect\citeauthoryear{{de Jong} et~al.,}{{de Jong} et~al.}{2015}]{DeJong2015}
{de Jong} J.~T.~A.,  et~al., 2015, \mn@doi [\aap] {10.1051/0004-6361/201526601}, \href {http://adsabs.harvard.edu/abs/2015A%26A...582A..62D} {582, A62}

\makeatother
\end{thebibliography}

\begin{appendix}

\section{Results Using Year 5 Magnitude Cuts and Photometric Redshifts}
\label{app:results_photz}
In the main text we assert that the use of Y1 magnitude cuts and true redshifts does not significantly affect our final conclusions (see Section \ref{sec:limitations}). Here, we describe in detail our results when using Y5 depth and photometric redshifts. Note that the DESC SRD only contains Y1 and Y10 cuts, so to obtain the Y5 magnitude cut we interpolate between the Y1 and Y10 cuts, using $\sqrt{\rm{time}}$ to scale the flux density, to obtain $i<24.92$.

\begin{figure*}
    \centering
    \includegraphics[width=\textwidth]{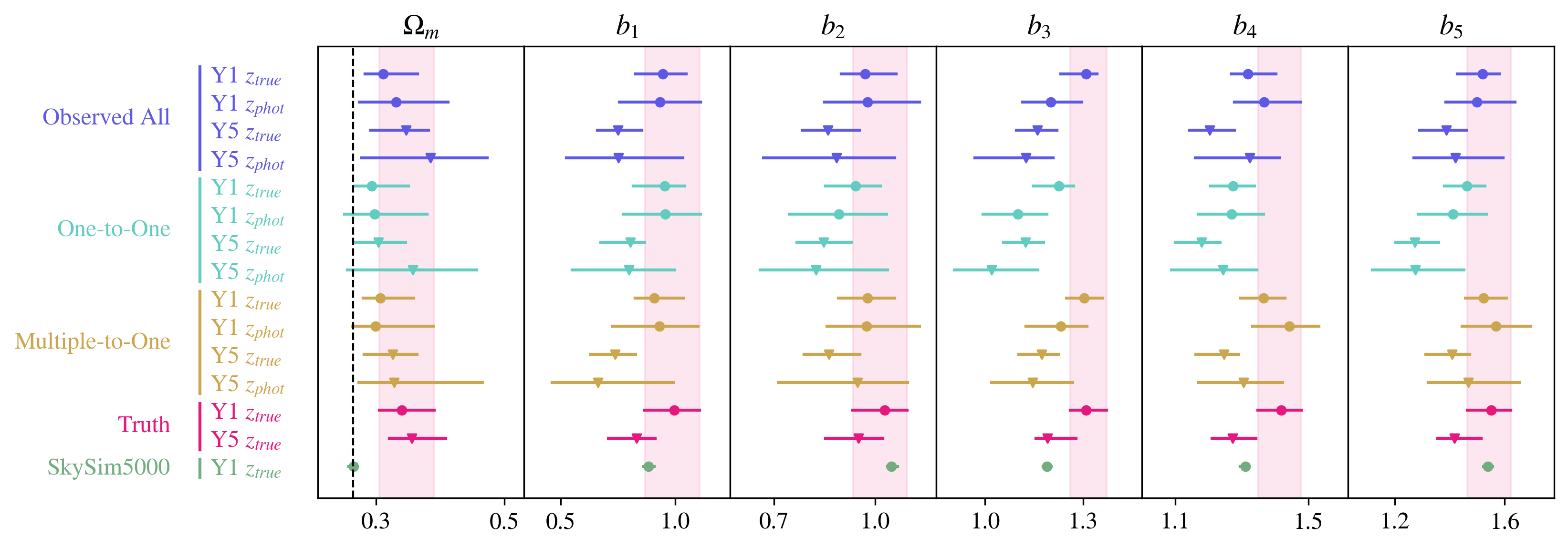}
    \caption{$1\sigma$ ranges for the inferred cosmological parameters of each sample, using different combinations of magnitude cutoff (Y1/Y5) and binning redshift (true redshift/photometric redshift). The first three samples (``observed all,'' ``one-to-one,'' and ``multiple-to-one'') are drawn from the DC2 catalog as described in Section \ref{ssec:matching}. The magenta ``truth'' sample is drawn from the CosmoDC2 truth catalog. The green sample in the bottom row corresponds to the SkySim5000 catalog, which we detail in Appendix \ref{app:skysim}. The simulation input value of $\Omega_{\rm m}=0.265$ is shown as a vertical dashed line. The magenta shaded regions show the $1\sigma$ range for the truth data in our fiducial (Y1) analysis.}
    \label{fig:year_photoz}
\end{figure*}

We exactly repeat our fiducial analysis using the LSST Y5 magnitude cut and a BPZ photometric redshift catalog for DC2. The latter can enter into the pipeline in two ways: first through the binning of galaxies when computing the correlation functions, and second through the $N(z)$ provided to \textsc{CCL}. For this test, we bin galaxies in the correlation functions according to their photometric redshifts (determined as the mean of the BPZ probability density function), and then feed the true $N(z)$ distribution for each of these photometric-redshift-selected bins into \textsc{CCL}. This corresponds to a scenario where we assume that we still have access to the true redshift distribution (i.e., we ignore any blending-related biases in the $N(z)$ distribution), since trying to estimate the impact of blending on $N(z)$ depends on the choice of $N(z)$ estimation method and is therefore beyond the scope of this paper. It is possible that blending-related errors in the calibrated $N(z)$ due to an imperfect estimator may lead to larger overall biases than we measure in this section. The $1\sigma$ confidence intervals we obtain are shown in Figure \ref{fig:year_photoz}.

Our main conclusions from the fiducial analysis are also apparent in these Y5 and photometric redshift tests. In all cases, the true value of $\Omega_{\rm m}$ is recovered within $<2\sigma$, and the fits to the observed data are consistent with the fits to the truth data within $1\sigma$. The inferred values of galaxy bias are slightly lower than the fiducial analysis for the Y5 tests by up to $1\sigma$. However, since these shifts are consistent between all three of our observed samples, they are unlikely to be caused by blending. We do not attempt further analysis to determine the source of the shifts in galaxy bias, but they are expected: using the Y5 magnitude cut introduces to the sample fainter galaxies that are less massive and less clustered, lowering the overall bias.

\begin{figure}
    \includegraphics[width=\columnwidth]{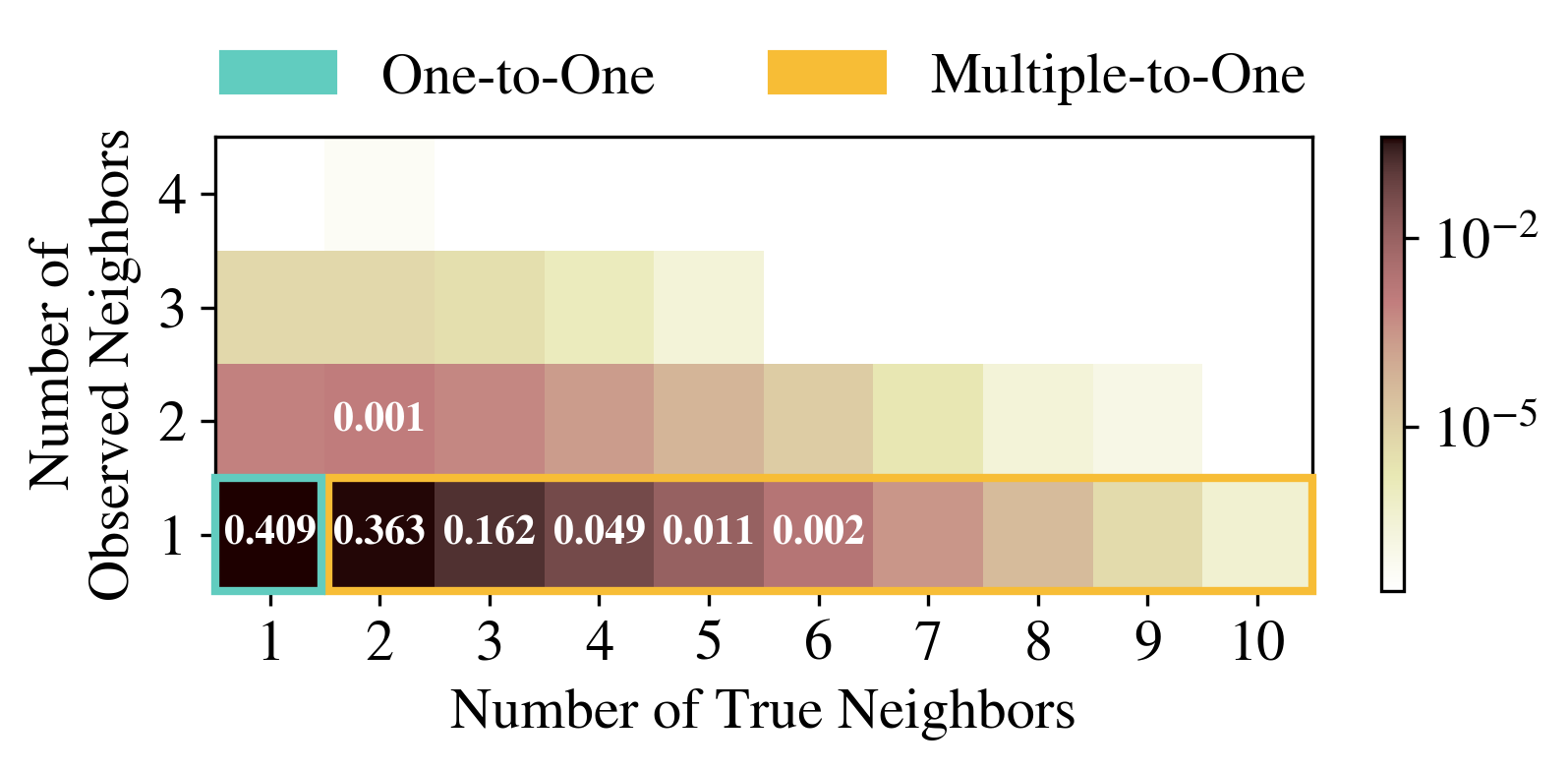}
    \caption{Distribution of neighbors in the truth and observed catalogs for each detected object, with a Y5 magnitude cut of $i<24.92$. Cells corresponding to our one-to-one and multiple-to-one samples are highlighted.}
    \label{fig:2dhist_y5}
\end{figure}

We now describe the characteristics of the blended samples in the Y5 and photometric redshift analyses and compare them to our main analysis. When using Y5 depth, the number of galaxies in our analysis between $0.2<z_{\rm true}<1.2$ nearly doubles from $13.1$ million to $22.1$ million. The fraction of blended objects remains approximately the same (Figure \ref{fig:2dhist_y5}), due to the fact that our definition of blends is independent of the sample magnitude cut and the blending rate is nearly uniform across magnitudes (Figure \ref{fig:mag_distribution}). The $N(z)$ distribution within the defined redshift bins is not significantly affected by the use of the Y5 cut. The qualities of the distribution are consistent with those we find in Y1: the multiple-to-one sample remains skewed to higher redshifts as compared to the one-to-one sample. In addition, the proportion of lost objects significantly increases (as expected, given that incompleteness has a larger effect for Y5).

\begin{figure}
    \includegraphics[width=\columnwidth]{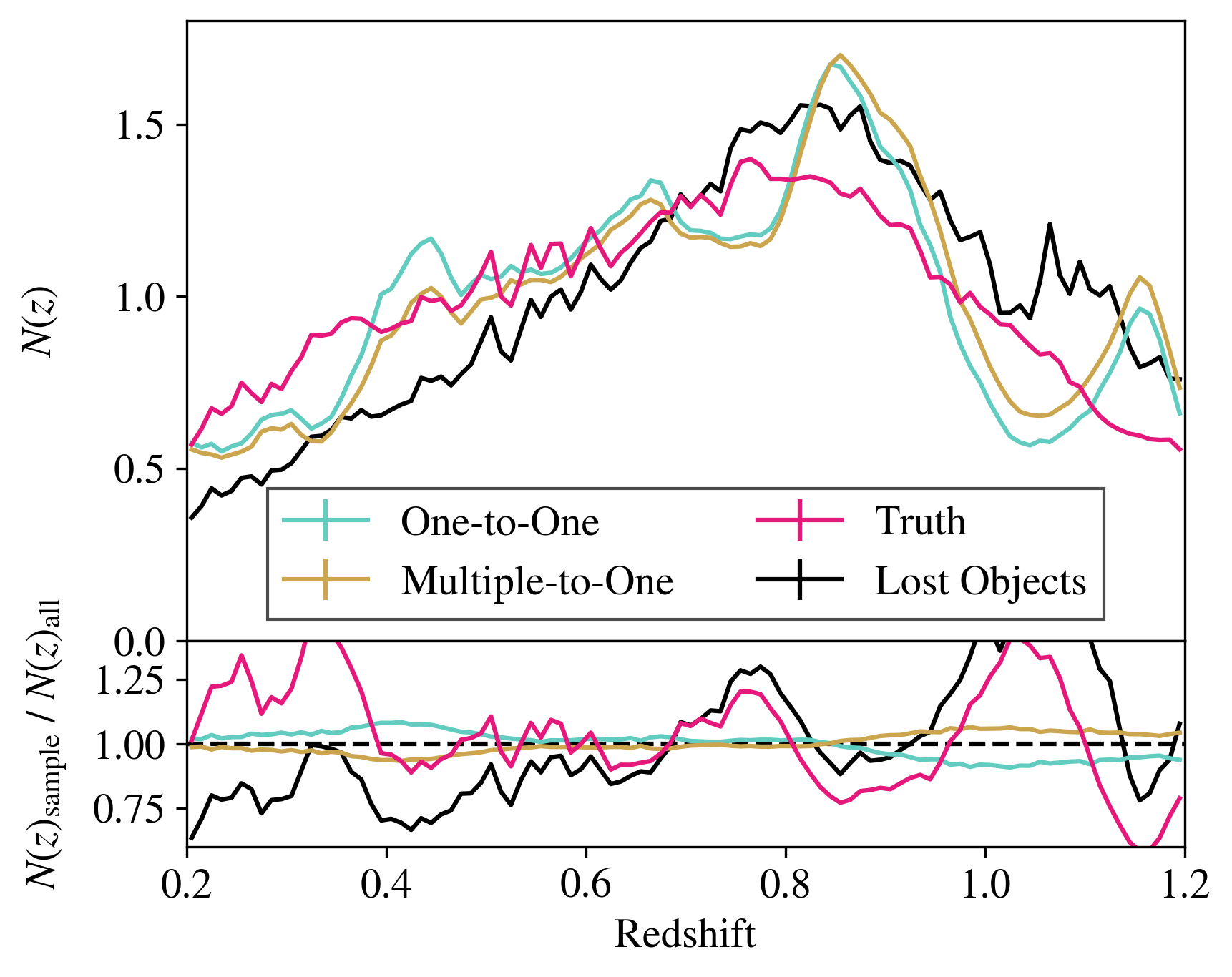}
    \caption{\emph{Top}: number of objects in our Y1 samples at a given redshift, normalized such that the integral over all bins is equal to unity. For the observed samples, the redshift is given by the photometric redshift. For the truth and lost samples, the redshift is given by the true redshift. \emph{Bottom}: deviation from the all observed sample.}
    \label{fig:Nz_Y1_photz}
\end{figure}

For the case of photometric redshifts, we use a catalog of redshifts estimated via BPZ using the measured photometry from the DC2 coadded images. Using these data, the overabundance at high redshifts of multiple-to-one objects as compared to one-to-one objects is reduced (Figure \ref{fig:Nz_Y1_photz}) as compared to the fiducial analysis (Figure \ref{fig:z_dist_all}). The $N(z)$ bias due to photometric redshift calibration with an isolated one-to-one sample is therefore below the $N(z)$ bias due to true (spectroscopic) calibration found in our fiducial analysis, so our main results provide an upper bound for the redshift bias in Y1. It is noteworthy that the one-to-one and multiple-to-one redshift distributions are similar when using photometric redshifts, given that every object in the multiple-to-one distribution has contaminated flux. Our results suggest that either the particular photometric redshift algorithm used here is robust to such contaminants, or that the majority of contaminants are too faint to significantly bias the measured redshift.

\begin{figure}
    \includegraphics[width=\columnwidth]{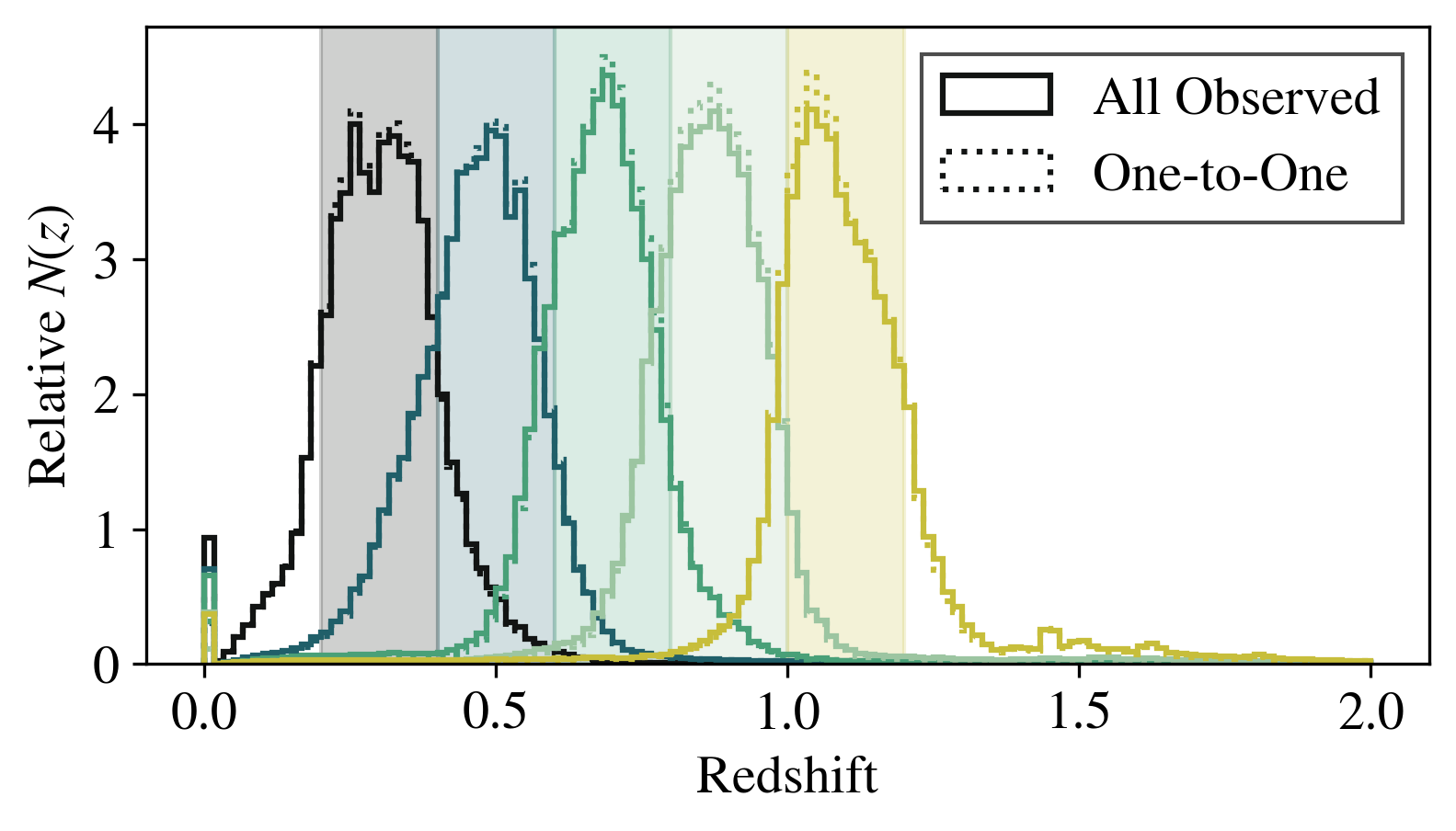}
    \caption{True redshift distribution of Y1-selected galaxies in each of our five tomographic bins, with bins assigned using photometric redshift. Each histogram is normalized such that the integral over all its bins is equal to unity. The shaded regions correspond to the selections for each tomographic bin. The overdensities at $z=0$ correspond to stars that are misidentified in the DC2 catalog and are not removed by our cuts.}
    \label{fig:Photoz_diagnostics_2}
\end{figure}

One feature of interest in Figure \ref{fig:Nz_Y1_photz} is the periodic offset between the true redshift distribution and the photometric distribution. This offset is a consequence of the photometric redshift algorithm and results in incorrect bin assignment when computing the correlation functions. This offset is consistent between the one-to-one and all observed samples. Figure \ref{fig:Photoz_diagnostics_2} shows the true redshift distribution of galaxies in each of our five tomographic bins, with bins assigned using photometric redshift. The true redshift distributions of the all observed and one-to-one samples are extremely similar. As shown in Figure \ref{fig:year_photoz}, incorrect bin assignment does not significantly impact the inferred value of $\Omega_{\rm m}$ for any of our samples.

All the results stated above are qualitatively the same when applying photometric redshifts to the Y5 catalog.

\section{Validation using the SkySim5000 cosmological simulation}
\label{app:skysim}

\begin{figure}
    \vspace{1em}
    \includegraphics[width=\columnwidth]{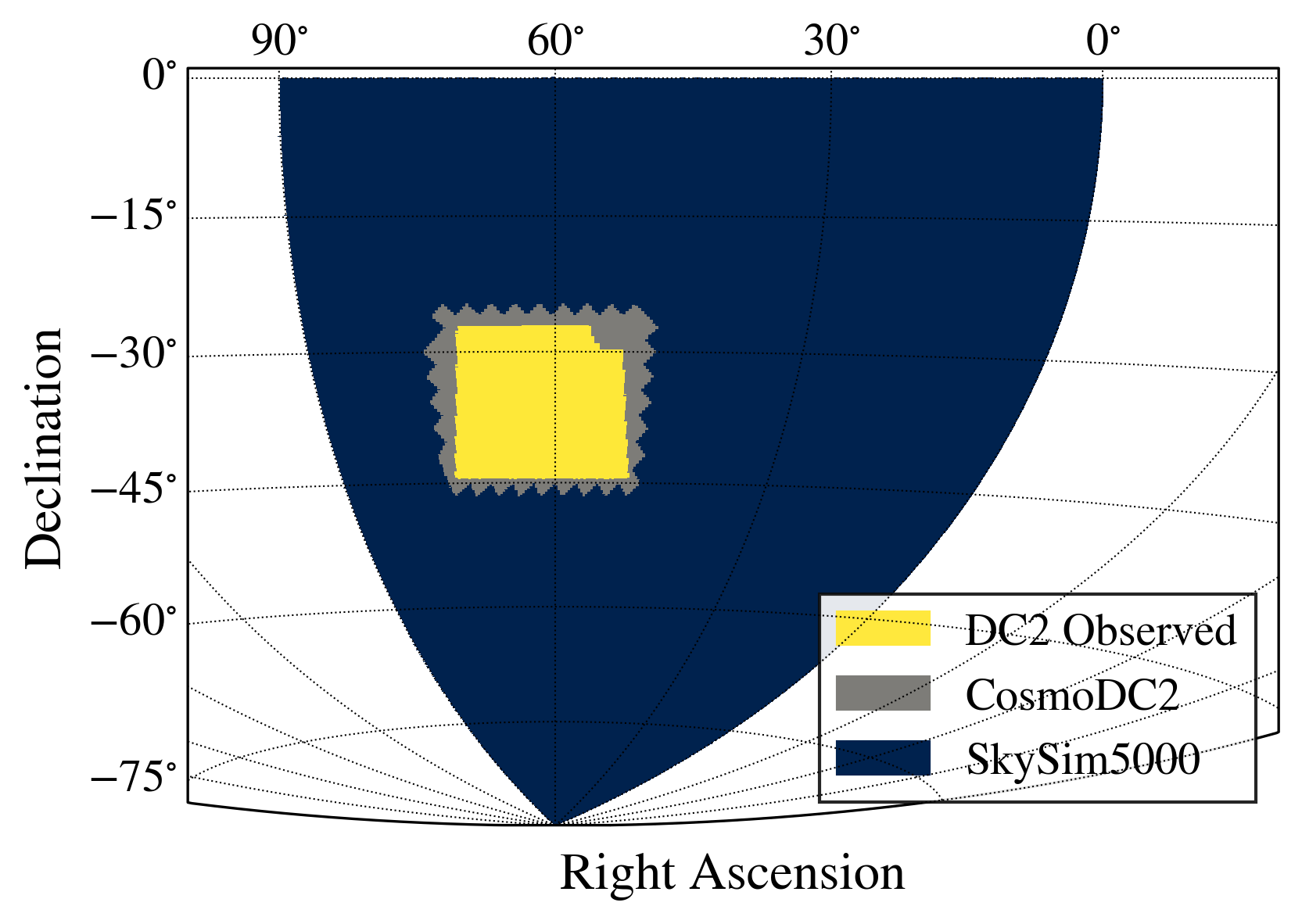}
    \caption{Footprints of the observed galaxies from the DC2 catalog (yellow), truth galaxies from the CosmoDC2 simulation (gray), and truth galaxies from SkySim5000 (blue). The footprints are shown using the Hammer-Aitoff projection.}
    \label{fig:skysim_map}
\end{figure}

The truth catalog used in this analysis (CosmoDC2) is a subset of the SkySim5000 simulated sky catalog, which has a footprint of one eighth of the sky. An illustration of CosmoDC2 and SkySim5000 (as well as the DC2 observed catalogs) is shown in Figure \ref{fig:skysim_map}. The two simulations use the same physics and differ only by their sizes, meaning that the measured correlation functions and inferred cosmological parameters should agree up to the level of statistical variance. We validate our measurements by comparing the aforementioned measurements in CosmoDC2 to those of \texttt{skysim5000\_v1.1.2}.

\begin{figure*}
    \includegraphics[width=2.1\columnwidth]{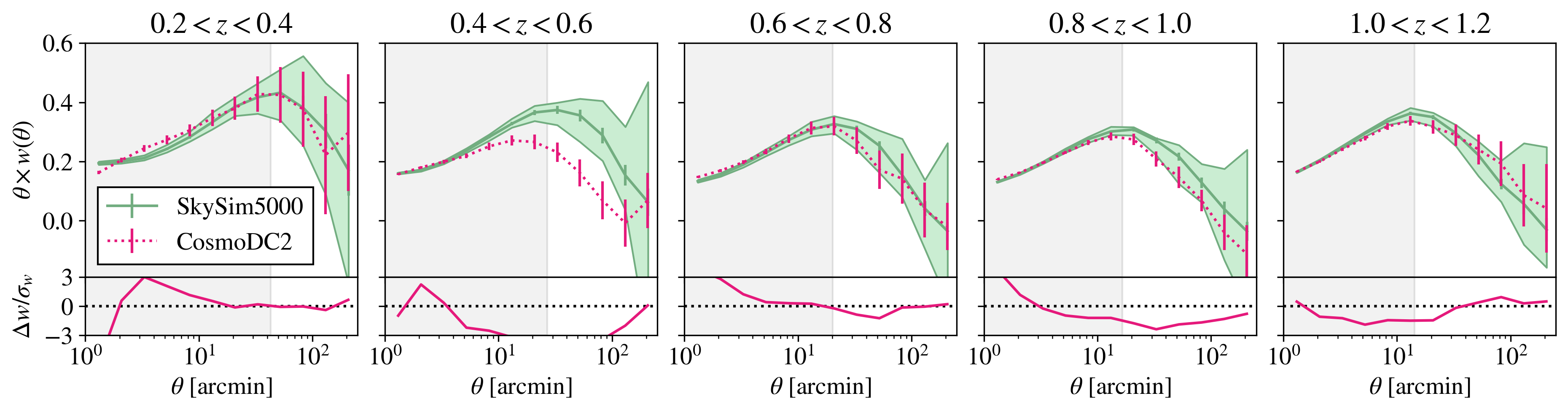}
    \caption{\emph{Top}: angular correlation function measurements for CosmoDC2 (magenta) and SkySim5000 (green). The $1\sigma$ distribution for 12 randomly-selected DC2-sized patches of SkySim5000 is shaded in light green. \emph{Bottom}: deviation from the full SkySim5000 correlation function. Gray shaded regions denote small scales ignored in cosmological analyses, as specified by the DESC SRD ($k<0.3h$\,Mpc$^{-1}$).}
    \label{fig:skysim_correlation}
\end{figure*}

The angular correlation function for galaxy clustering is shown in Figure \ref{fig:skysim_correlation}. The CosmoDC2 measurements are statistically compatible within $1\sigma$ of SkySim5000 for all but the second redshift bin, which differs significantly. In order to confirm that the discrepancy is likely due to cosmic variance, we randomly select 12 patches of sky from SkySim5000 with areas matching that of CosmoDC2. The $1\sigma$ distribution of correlation functions from these patches is plotted in Figure \ref{fig:skysim_correlation}. CosmoDC2 lies within $2-3\sigma$ of the patches for the second redshift bin, and well within $1\sigma$ for the other bins. It is likely that the deviations in CosmoDC2 are simply an effect of cosmic variance, rather than a systematic issue with the catalog.

\begin{figure*}
    \centering
    \includegraphics[width=.89\textwidth]{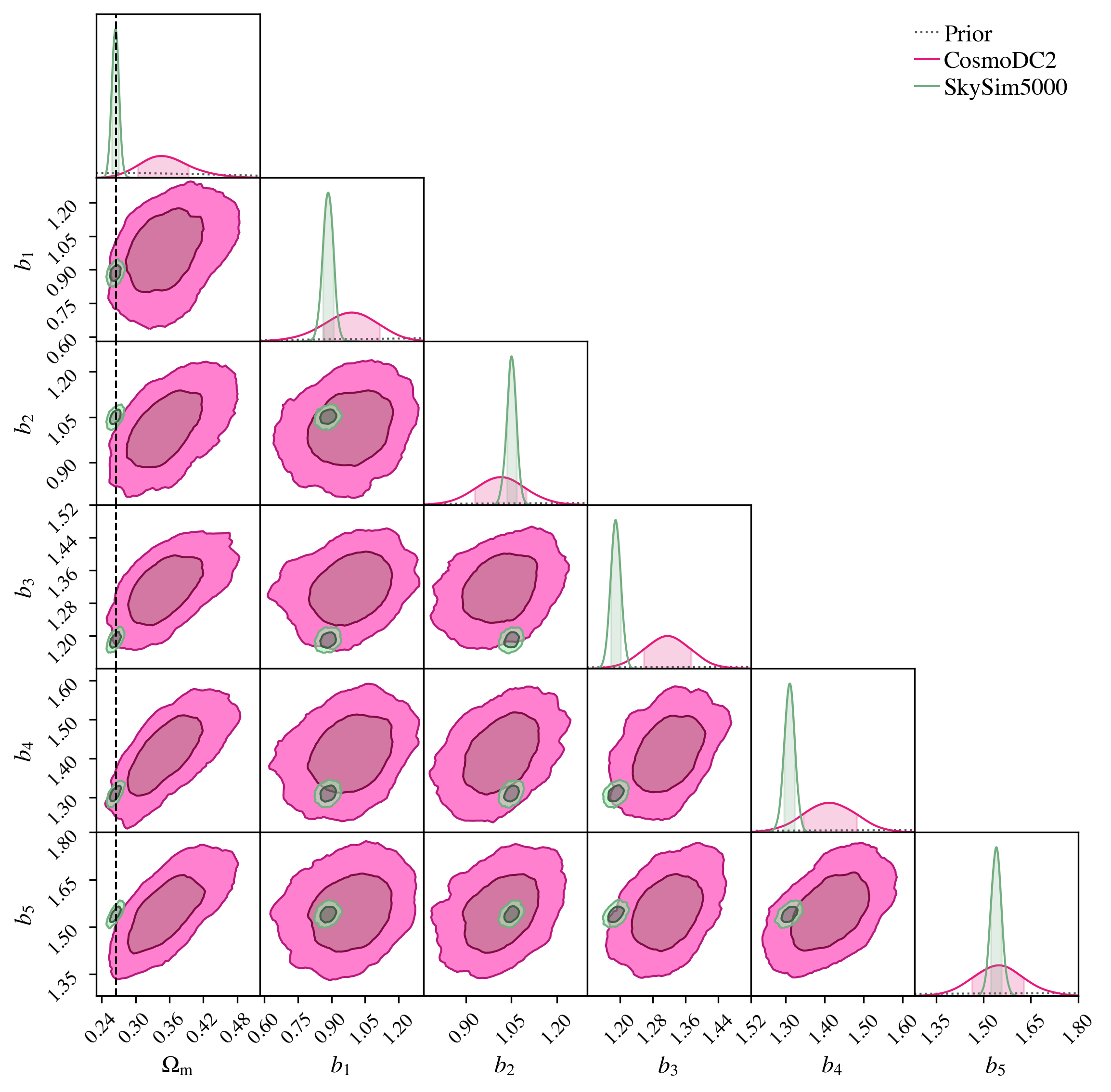}
    \caption{Contours for cosmological parameters assuming $\Lambda$CDM model with $\Omega_{\rm m}$ and galaxy bias $b_{i}$ freed, computed for CosmoDC2 (magenta) and SkySim5000 (green). We use DESC SRD priors and show the simulation input value of $\Omega_{\rm m}=0.265$ as a vertical dashed line.}
    \label{fig:posteriors_skysim}
\end{figure*}

We repeat the MCMC analysis described in Section \ref{ssec:cosmology} for SkySim5000 and show the results in Figure \ref{fig:posteriors_skysim}. The contours for SkySim5000 are more precise than for CosmoDC2 thanks to the greater statistical power, and we recover the simulation input $\Omega_{\rm m}$ within $1\sigma$. The contours agree with those of CosmoDC2 within $2-3\sigma$, resembling the level of statistical deviation in the correlation function's second redshift bin.

\end{appendix}

\end{document}